\documentclass[aos,preprint]{imsart}

\RequirePackage[OT1]{fontenc}
\RequirePackage{amsthm,amsmath,natbib}
\RequirePackage[colorlinks]{hyperref}
\usepackage{array}
\usepackage{mathrsfs}
\usepackage{amssymb,bbm}
\usepackage{amsfonts}
\usepackage{amsmath,amssymb}
\usepackage{graphicx}
\usepackage{amscd}
\usepackage{color}
\usepackage{mathrsfs}
\usepackage{latexsym}
\usepackage{array,bm}
\usepackage{epstopdf}
\usepackage{enumerate}
\usepackage{subcaption}
\usepackage{soul}
\usepackage{ulem}

\usepackage{chngcntr}

\usepackage[flushleft]{threeparttable}

\usepackage{textcomp}
\usepackage{multicol}
\usepackage{algorithm}
\usepackage{algorithmic}
\usepackage{booktabs}

\def\boxit#1{\vbox{\hrule\hbox{\vrule\kern6pt
          \vbox{\kern6pt#1\kern6pt}\kern6pt\vrule}\hrule}}

\newcommand{\vF}{\bm{F}}

\newcommand{\btheta}{\bm{\theta}}

\newcommand{\bW}{\bm{W}}
\newcommand{\bomega}{\bm{\omega}}

\newcommand{\bb}{\bm{b}}

\newcommand{\bA}{\bm{A}}

\newcommand{\argmin}{\mathop{\mathrm{argmin}}}

\allowdisplaybreaks[1]

\startlocaldefs
\theoremstyle{plain}
\endlocaldefs

\definecolor{gray}{gray}{0.5}

\begin{document}

\begin{frontmatter}
\title{Enhancement of price trend trading strategies via image-induced importance weights}\thankstext{t2}{The codes and data used for this paper are accessible at \url{https://github.com/ZhuZhouFan/TWMA}.}

%\vspace{20mm}

\begin{aug}

  \author[a]{\fnms{Zhoufan} \snm{Zhu}
  \ead[label=e1]{tylerzzf1103@gmail.com}}
  \and
  \author[b]{\fnms{Ke} \snm{Zhu} %\thanksref{t2}
  \ead[label=e3]{mazhuke@hku.hk}}

 \affiliation[a]{Wang Yanan Institute for Studies in Economics (WISE), Xiamen University, China}
 \affiliation[b]{Department of Statistics and Actuarial Science, University of Hong Kong, Hong Kong}

\end{aug}

\vspace{3mm}

%%%%%%%%%%%%%%%%%%%%%%%%%%%%%%%%%%%%%%%%%%%%%%%%%%%%%%%%%%%%%%%%%%%%%%%%%%%%%%%%%%%%%%%%%%%%%%%%%%%%%%%%%%%%%%%%%%%%%%%%%%%
%
%                          Abstract
%
%%%%%%%%%%%%%%%%%%%%%%%%%%%%%%%%%%%%%%%%%%%%%%%%%%%%%%%%%%%%%%%%%%%%%%%%%%%%%%%%%%%%%%%%%%%%%%%%%%%%%%%%%%%%%%%%%%%%%%%%%%%
\begin{abstract}

We open up the ``black-box'' to identify the predictive general price patterns in price chart images via the deep learning image analysis techniques. Our identified price patterns lead to the construction of image-induced importance (triple-I) weights, which are applied to weighted moving average the existing price trend trading signals according to their level of importance in predicting price movements. From an extensive empirical analysis on the Chinese stock market, we show that the triple-I weighting scheme can significantly enhance the price trend trading signals for proposing portfolios, with a thoughtful robustness study in terms of network specifications, image structures, and stock sizes. Moreover, we demonstrate that the triple-I weighting scheme is able to propose long-term portfolios from a time-scale transfer learning, enhance the news-based trading strategies through a non-technical transfer learning, and increase the overall strength of numerous trading rules for portfolio selection.

%Although deep learning models have achieved impressive success in automatically learning predictive price patterns from extensive stock price images, two critical questions remain unresolved for both researchers and practitioners. One is what general price patterns do deep learning models learn from price charts, the other one is how can traders leverage these detected patterns to enhance their own trading strategies? To address these questions, this paper proposes the \underline{t}riple-I \underline{w}eighted \underline{m}oving \underline{a}verage (TWMA) method, offering novel perspectives to encourage further in-depth studies. The TWMA method first develops a new artificial intelligence ``trader,'' equipped with the \underline{res}idual \underline{net}work (ResNet), to forecast the price trends. Next, the TWMA method derives \underline{i}mage-\underline{i}nduced \underline{i}mportance (triple-I) weights, which reflect the importance levels of all time points within the price image for the ResNet ``trader'' in predicting future price movements. Finally, the TWMA method applies these weights to weighted moving average the signals produced by the trader's own strategies, thereby enhancing trading strategies through a new signal adjusted according to the significance of historical time points. An application to a battery of technical and non-technical signals using the Chinese A-share stock market dataset demonstrates the effectiveness of the TWMA method, particularly in enhancing trading strategies for large-cap stocks.
\end{abstract}

%\begin{keyword}[class=MSC]
%\kwd[] {C13,} {C22,} {C58.}
%\end{keyword}

\begin{keyword}
\kwd{AI in finance; Image-induced importance weights; Portfolio selection; ResNet ``trader''; Technical analysis; Trading strategy enhancement} %\kwd{\LaTeXe}
\end{keyword}

\end{frontmatter}

%%%%%%%%%%%%%%%%%%%%%%%%%%%%%%%%%%%%%%%%%%%%%%%%%%%%%%%%%%%%%%%%%%%%%%%%%%%%%%%%%%%%%%%%%%%%%%%%%%%%%%%%%%%%%%%%%%%%%%%%%%%%%%%%%%%%%%
%
%                                          Section 1  Introduction
%
%%%%%%%%%%%%%%%%%%%%%%%%%%%%%%%%%%%%%%%%%%%%%%%%%%%%%%%%%%%%%%%%%%%%%%%%%%%%%%%%%%%%%%%%%%%%%%%%%%%%%%%%%%%%%%%%%%%%%%%%%%%%%%%%%%%%%%
\newpage

\setcounter{equation}{0}

\section{Introduction}\label{sec:introduction}

Technical analysis (also known as ``charting'') is widely used by investors to forecast future price trend of stocks,
based on the trading signal observed from price charts that consist of historical stock price and volume data. Although it cannot unequivocally conclude the reliability and effectiveness of technical analysis, an extensive literature spanning the past decades has explored the usefulness of technical analysis from both theoretical and empirical perspectives. See, for example, \cite{David1989tech}, \cite{Bruce1989trade}, \cite{Lawrence1994market}, \cite{BARBERIS1998307}, and \cite{han2016trend} for the existence of equilibrium predictability with technical analysis, and \cite{brock1992simple}, \cite{lo2000foundations}, \cite{zhu2009technical}, \cite{neely2014forecasting}, \cite{detzel2021learning}, and \cite{murray2024charting} for strong empirical support for technical trading rules.

%BarberisPsychology2018

Till now, the standard trading strategies in technical analysis root in two widely documented stock price patterns: reversal \citep{reversal, lo2015reversal} and momentum \citep{momentum}, which are predictive for stock returns but challenging to explain using typical factor models (\citealp{han2016trend}). However, stock price dynamics are influenced by human behavior and psychology (\citealp{barberis2018psychology}), leading to certain subtle and complex price patterns that could not be efficiently captured by these standard trading strategies. To solve this issue, \cite{jiang2023re} apply the machine learning tool of convolutional neural network (CNN) to propose a trading strategy, using image-based CNN predictions for the direction (up or down) of stock returns. To mimic investor perceptions and decision processes in real world,
their strategy creates a CNN ``trader'', who can read general price patterns from price chart images to forecast the direction of future stock returns for selecting portfolios. As demonstrated by the empirical studies in \cite{jiang2023re},
the performance of this CNN ``trader'' is encouraging with certain insightful interpretations. The portfolio strategy made by the CNN ``trader'' outperforms those using the standard reversal and momentum trading strategies, which can only explain a small part of trading signals extracted by the CNN ``trader''. More importantly, the general price patterns detected by the CNN ``trader'' appear to be context-independent, since they could hold in different markets or at different time scales according to a thoughtful transfer learning investigation.

Although the CNN ``trader'' is empirically successful, he overlooks two important questions that are practically interesting to the traders in real world: First, what are the general price patterns detected from price charts? Second, how can the traders make use of these detected general price patterns to enhance their own trading strategies? In this paper, we contribute to the literature by partly tackling the above two questions. For the first question, we design a new artificial intelligence ``trader'', who is equipped with the residual network (ResNet) in \cite{he2016resnet}, to forecast the direction of future stock returns. This ResNet ``trader'' is potentially smarter than the CNN ``trader'', since he adopts much deeper neural networks without suffering from the vanishing/exploring gradient problem to extract trading signals from the price chart images. Moreover, this ResNet ``trader'' offers us two visual localization maps of each price chart image via the smooth gradient-weighted class activation mapping (smooth Grad-CAM) method in \cite{omeiza2019smooth}. Each spatial location of the localization map is assigned with a value of importance, which represents the importance level of the extracted information at this spatial location for predictions. The ``upward'' localization map identifies important regions of price chart image (with larger values of importance) for predicting an upward future stock price movement, whereas the ``downward'' localization map does it for predicting a downward future stock price movement. Since these important regions contain more useful information for the direction prediction of stock return, they are more likely to reveal the general price patterns detected by the ResNet ``trader''.

%, and \cite{murray2024charting}

For the second question, we propose a novel weighted moving average method, aiming to enhance the performance of existing price trend trading strategies. Our method is based on the image-induced importance (III, or so-called triple-I) weights. For each price chart image, we have two sequences of triple-I weights, computed from two localization maps. Specifically, the triple-I weights from the ``upward'' (or ``downward'') localization map are averaged values of importance vertically, so their values after resizing and normalization reflect the importance level of all time points in the price chart image for predicting an upward (or downward) future stock price movement. Notably, the ResNet ``trader'' can provide the predicted value of ``upward'' probability of future stock price movement, guiding us to decide which sequence of triple-I weights to utilize. Once the triple-I weights are determined, our method applies them to weighted moving average the signals produced by pre-specific price patterns (e.g., momentum) or trading rules (e.g., filter rules), leading to new trading strategies.
The idea behind the triple-I weighted moving average (TWMA) method above is logical: It aims to amplify (or diminish) the
trading signal at a more (or less) important time point according to its value of triple-I weight. The TWMA method appears to be the first prototype that utilizes signals extracted from price chart images by a machine to teach traders how to enhance their technical analysis. Undoubtedly, its performance offers us evidences for validating whether the ResNet ``trader'' is capable of extracting valuable information from price chart images.

Our empirical analysis demonstrates the practical performance of the TWMA method in portfolio selections for the largest 1,800 stocks in the Chinese A-share stock market from January 1, 2014 to May 1, 2023. The implementation of TWMA method centers around the computation of triple-I weights. Based on the input of price chart images that consist of price and volume data over the past 5, 20, or 60 days, the ResNet ``trader'' trains a ResNet with 18, 34, or 50 layers to predict the ``downward'' and ``upward'' probabilities of future stock price movements over a forward horizon of $1$ or 5 days. Using a trained ResNet, the ResNet ``trader'' computes the triple-I weights from the ``downward'' and ``upward'' localization maps to weighted moving average each of five commonly used price trend technical trading signals: 2-12 momentum (MOM), 1-month short-term reversal (STR), 1-week reversal (WSTR), trend (TREND, \citealp{han2016trend}), and formulaic alphas (ALPHA, \citealp{kakushadze2016alpha}), leading to five different TWMA-based signals. In terms of out-of-sample trading strategies, we sort stocks by their predicted values of each considered trading signal to propose the equal-weight long-short decile portfolios, whose holding period is equal to the forward horizon.

Our primary findings from the empirical analysis are three-fold. First, we find that the portfolios selected by each TWMA-based signal attain substantially higher values of annualized Sharpe ratio and smaller values of monthly turnover than those selected by the corresponding original trading signal. Specifically, our best one-day portfolios are constructed under the TWMA-based trading signals from a trained ResNet with 34 layers using the input of 5-day price chart images.
Their values of Sharpe ratio under the TWMA-based MOM, STR, and WSTR signals are at least 219\% higher than those under the original MOM, STR, and WSTR signals. This advantage remains when considering the TREND and ALPHA signals, with the Sharpe ratio values increased by 27\% and 73\%, respectively, after the TWMA-based signal enhancement. For the purpose of comparison, when the triple-I weights are computed by a trained CNN (from the CNN ``trader''), the performance of TWMA-based portfolios generally becomes worse, shedding light on the necessity of deeper neural networks (e.g., the ResNet) to extract information from price chart images. Unlike one-day portfolio selections, the one-week portfolios based on the TWMA-based signals do not show a remarkable advantage over those based on the original signals, although in general the former has smaller values of turnover than the latter.
This finding is consistent with that in \cite{jiang2023re}, indicating the difficulty to grab informative general price patterns from price chart images for making longer-term return predictions. Furthermore, we conduct a series of robustness checks in terms of the structure of image inputs and the size of stocks, lending the support to above conclusions.

Second, we find the capacity of the TWMA-based method on image-based transfer learning from two perspectives: time-scale transfer learning and non-technical transfer learning. The time-scale transfer learning involves using the triple-I weights determined by the high-frequency ResNet ``trader'' to weighted moving average the technical trading signals unfolded at a lower frequency, thereby enhancing price trend strategies with a longer holding period. For an illustration, we chart the 20-day price-volume histories into a 5-window condensed image, with each window being a 4-day interval. We then transfer the ResNet ``trader'' trained for analyzing high-frequency images to detect the critical regions in these condensed images, further enhancing the technical trading signals in one-month (i.e., 20-day) portfolio selections without re-training. The transferred triple-I weights improve all considered five trading signals in one-month portfolio selections by increasing Sharpe ratio values from $0.16$, $0.37$, $0.98$, $0.85$, and $0.73$ to $0.21$, $0.52$, $1.05$, $0.88$, and $0.79$, respectively. For the non-technical transfer learning, we apply the triple-I weights computed from price chart images to enhance the selection of portfolios via the news signals \citep{Zhou2024Text}. The news signal is non-technical and it is the sentiment score drawn from news articles. Intuitively, the crucial events reported by influential news articles should be reflected as an impulse on price trends and then be detected by the ResNet ``trader'' from price chart images. Hence, we expect that using the triple-I weights can help de-noise the sentiment scores extracted from noisy news articles. Encouragingly, our empirical results match this expectation: We find that for one-day and one-week portfolio selections,
the TWMA-based news signals using a trained ResNet earn Sharpe ratio values of $2.04$ and $0.41$, respectively, whereas the original news signals only attain Sharpe ratio values of $1.72$ and $0.15$, respectively. In this sense, our non-technical transfer learning based on the TWMA method is interesting in its own rights, since it offers a novel way to transfer knowledge from price-volume data to financial text data for trading enhancement.

Third, we find the overall function of the TWMA-based method to enhance $7,846$ distinct trading rule signals \citep{sullivan1999data, BAJGROWICZ2012473} for portfolio selections. For each trading rule signal, we derive its TWMA-based counterpart to propose one-day and one-week portfolios. Regarding one-day portfolios, the TWMA-based trading rule signals using the trained ResNet yield a distribution of $7,846$ Sharpe ratio values with the sample mean of $0.25$; in contrast, the original trading rule signals (or the TWMA-based trading rule signals using the trained CNN) deliver a Sharpe ratio value distribution with the sample mean of $-0.10$ (or $-0.06$). Moreover, based on the the rule-by-rule differenced Sharpe ratio values between any two kinds of signals, we obtain the statistical evidence from Student's $t$-test that the TWMA-based trading rule signals using the trained ResNet have the significant advantage over their two competitors in one-day portfolio selections. Notably, we can also observe similar phenomena in one-week portfolio selections.

The remaining paper is organized as follows. Section \ref{sec:method} introduces our ResNet ``trader'', with the used network architecture, training procedure, and implementation details. Section \ref{sec:TWMA} raises the localization maps and proposes
the TWMA method. Section \ref{sec:empirical} provides our empirical studies on enhancing price trend trading strategies in the Chinese A-share stock market via the TWMA method. Sections \ref{sec:time_transfer} and \ref{sec:news_rules} explore the capacity of TWMA method in time-scale transfer learning and non-technical transfer learning, respectively. Section \ref{trading_rules} analyzes the benefits of TWMA method in the context of trading rules. Concluding remarks are offered in Section \ref{sec:conclusion}. Some preliminary knowledge on the ResNet, a computational algorithm, and additional empirical results are reported in the Appendices.

\section{The ResNet ``Trader''}\label{sec:method}

Analyzing price charts of financial assets is a routine work for many traders in the market. In this section, we create a ResNet ``trader'', who mimics the behavior of traders by exploring price patterns from price chart images.

\subsection{The Design of Price Chart Images} \label{sec:image}
Each image used by our ResNet ``trader'' includes the price and volume information from a price chart over consecutive $D$ days. It is constructed via the ``\href{https://github.com/matplotlib/mplfinance}{mplfinance}'' module in Python to depict the open-high-low-close (OHLC) bars, volume bars, and a $D$-day moving average price line, which, as explained in \cite{jiang2023re}, can convey a variety of information altogether on price trends, volatility, intraday and overnight return patterns, and trading volume.

\begin{figure}[!htp]
    \centering
    \includegraphics[width = 0.8\linewidth]{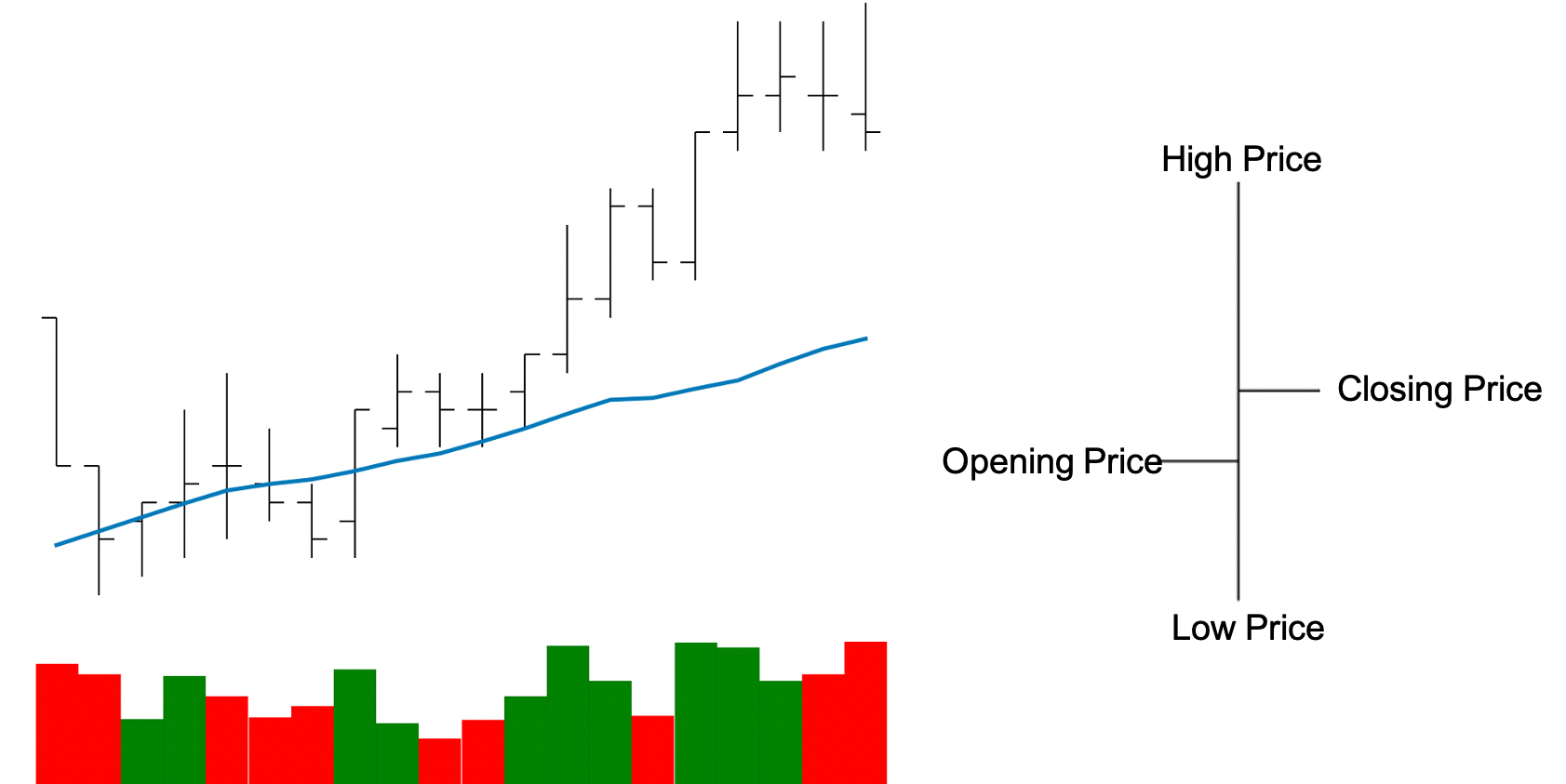}
    \caption{(Color online) Left panel: a 20-day image with OHLC bars (in black), volume bars (in green or red), and a 20-day moving average price line (in blue), based on the daily market data for Bank of China from March 24, 2023 to April 21, 2023. Right panel: an OHLC bar indicating daily opening, high, low, and closing prices.}
    \label{fig:ohlc}
\end{figure}

Specifically, we set the height of each image to be 224 pixels, and then display the OHLC bars (including the moving average price line) on the top 78\% area of the image, exhibit the volume bars in the bottom 17\% area of the image, and use 5\% empty space of the image in between. For each daily OHLC bar, it has a vertical line (one pixel wide) with its top and bottom to represent the high and low prices, respectively; meanwhile, it also has a line to the left (five pixels wide) for marking the opening price and a line to the right (five pixels wide) for marking the closing price. For two consecutive daily OHLC bars, there is a gap of five pixels to ease the visualization.
The moving average price line is drawn by connecting all of $D$-day moving average prices, each of which is a dot with one pixel wide. All of daily OHLC bars and the moving average price line are re-scaled in a way that the maximum and minimum prices (including both OHLC and moving average prices) coincide with the top and bottom of the top 78\% area of the image. For each daily volume bar, its height represents the trading volume, while its width spans eleven pixels wide. Similarly, all of daily volume bars are re-scaled such that their maximum height aligns with the height of the bottom 17\% area of the image. In terms of visual appeal, we use a white background color in the image, set the color of OHLC bars and moving average price line to be black and blue for preventing confusion caused by the crossing, and choose the color of green and red for volume bars to present ``up'' and ``down'' days as done in Yahoo Finance. By construction, our image has the size $224\times 11D$ pixels. To match the conventional size of image imputed into the ResNet, we resize our image to $224 \times 224$ pixels using the classical bilinear interpolation method.
In the sequel, our proposed image is called $D$-day price chart image (or $D$-day image for brevity).
As an illustrating example, one of 20-day images for Bank of China from March 24, 2023 to April 21, 2023 is presented in Fig\,\ref{fig:ohlc}.

\subsection{The ResNet Learning Procedure}

Like most of traders in real world, the ResNet ``trader'' aims to recognize informative trading signal from geometric shapes in the images.
The differential is that the traders do it manually using human cognition, whereas the ResNet ``trader'' does it automatically via the machine learning tool of ResNet \citep{he2016resnet}.

The input of ResNet model is a $D$-day image.
Note that a $D$-day image for stock $i$ on day $t$ has the size $224 \times 224$ pixels. Since each pixel is determined by a $3\times 1$ vector (RGB intensities), this $D$-day image can be represented by a three-dimensional tensor $\vF_{i,t,D} \in \mathcal{R}^{224 \times 224 \times 3}$.

 Next, the ResNet model applies $L$ blocks to extract the general price patterns from a $D$-day image represented by $\vF_{i,t,D}$, where $L$ is the depth of the ResNet model, and three operations: convolution, activation, and max-pooling, are implemented successively in each block. Convolution can be considered a specific type of kernel smoothing, scanning the image horizontally and vertically to generate a summary of the image contents surrounding each position. Activation applies a nonlinear  transformation, such as the rectified linear unit function (ReLU), on the output of convolution in an element-wise manner. Max-pooling, the final operation in the block, scans the output of activation and returns the highest value over the surrounding areas in the image to reduce data dimensionality and noise. Unlike convolution and activation, max-pooling is unnecessary and sometimes not implemented in the block. Below, we let $\mathrm{Conv}(\cdot; \btheta)$ denote the convolution operation depending on the parameter vector $\btheta$, $\mathrm{ReLU}(\cdot)$ be an entry-wise vector-valued function used for activation (i.e., non-linear transformation), and $\mathrm{MaxPool}(\cdot)$ denote the max-pooling operation. For the sake of brevity, we defer more details on the computation of these three operations into the Appendix A.

Third, we introduce the architecture of $L$ blocks in the ResNet model. Given the input $\vF_{i,t}^{(0)} \equiv \vF_{i,t,D}$,
the output formula for the first block is as follows:
\begin{align}\label{res_1}
    \vF_{i,t}^{(1)} = \mathrm{MaxPool}\left\{\mathrm{ReLU}\left[\mathrm{Conv}\left(\vF_{i,t}^{(0)}; \btheta^{(1)}\right)\right]\right\} \in \mathcal{R}^{H^{(1)} \times W^{(1)} \times C^{(1)}},
\end{align}
where $\vF_{i,t}^{(1)}$ is the extracted feature map by the first block, and $H^{(1)}$, $W^{(1)}$, and $C^{(1)}$ are positive integers representing the height, width, and number of channels of $\vF_{i,t}^{(1)}$, respectively. For the remaining $L-1$ blocks, the output formula for the $l$-th block is computed recursively as follows:
\begin{align}\label{res}
\begin{split}
    \vF_{i,t}^{(l)}  = \mathrm{ReLU}\left[\mathrm{Conv}\left(\vF_{i,t}^{(l - 1)}; \btheta^{(l)}\right)\right]
     + \mathbb{R} \left(\vF_{i,t}^{(l - S)} \right) \mbox{I} (l = S \times n +1)\in \mathcal{R}^{H^{(l)} \times W^{(l)} \times C^{(l)}}
\end{split}
\end{align}
for $l=2, \cdots, L$, where $\vF_{i,t}^{(l)}$ is the extracted feature map by $l$-th block,  $\mathbb{R}(\cdot)$ is the rescale operator for ensuring those two terms in the right-hand side have the same shape, $S$ is a user-specific positive integer, $\mbox{I}(\cdot)$ is the indicator function with a value of one only when $l = S \times n +1$ for a positive integer $n$, and $H^{(l)}$, $W^{(l)}$, and $C^{(l)}$ are positive integers representing the height, width, and number of channels of $\vF_{i,t}^{(l)}$, respectively. The term $\mathbb{R} \big(\vF_{i,t}^{(l - S)} \big)$ is the key component in formula \eqref{res} to locate the information extracted $S$ blocks before. Due to the presence of $\mbox{I} (l = S \times n +1)$, this key component affects $\vF_{i,t}^{(l)}$ directly via (\ref{res}) only when there exists a positive integer $n$ such that $l=S\times n+1$. For example, when $S=2$ and $l=3$, $\mathbb{R} \big(\vF_{i,t}^{(1)} \big)$ can directly affect $\vF_{i,t}^{(3)}$, since $l=S\times 1+1$; however, when $S=2$ and $l=4$, $\mathbb{R} \big(\vF_{i,t}^{(2)} \big)$ cannot directly affect $\vF_{i,t}^{(4)}$, since $l\not=S\times n+1$ for any positive integer $n$. Re-writing formula \eqref{res} as
\begin{align}\label{res_rewrite}
\begin{split}
    \vF_{i,t}^{(l)} - \mathbb{R} \left(\vF_{i,t}^{(l - S)} \right) \mbox{I} (l = S \times n +1) & = \mathrm{ReLU}\left[\mathrm{Conv}\left(\vF_{i,t}^{(l - 1)}; \btheta^{(l)}\right)\right],
\end{split}
\end{align}
it is evident that the right-hand side of (\ref{res_rewrite}) captures the residual between the higher-level feature map $\vF_{i,t}^{(l)}$ and its lower-level counterpart $\vF_{i,t}^{(l - S)}$ (after re-scaling), provided that $\mbox{I} (l = S \times n +1)=1$. Hence, formula \eqref{res} is called the residual learning.

After recursively extracting patterns through $L$ blocks to obtain the final feature map $\vF_{i,t}^{(L)}$, the ResNet model applies the average-pooling operation (denoted by $\mathrm{AvgPool}(\cdot)$) to further de-noise $\vF_{i,t}^{(L)}$. Specifically, this average-pooling operation calculates the mean of all pixel values in each channel of $\vF_{i,t}^{(L)}$ to produce channel-wise scores $\bar{\vF}_{i,t}$:
\begin{align}\label{avg_pool}
\begin{split}
     & \bar{\vF}_{i,t}  = \mathrm{AvgPool}(\vF^{(L)}_{i,t}) = \left(\bar{F}_{i,t}(1), \dots, \bar{F}_{i,t}\left(C^{(L)} \right) \right)' \in \mathcal{R}^{C^{(L)}}
    \\
    & \text{with} \,\,\, \bar{F}_{i,t}(c)  = \frac{1}{H^{(L)}W^{(L)}} \sum_{h = 1}^{H^{(L)}} \sum_{w = 1}^{W^{(L)}} F^{(L)}_{i,t}(h,w,c),
\end{split}
\end{align}
where $F^{(L)}_{i,t}(h,w,c)$ is the $(h,w,c)$-th entry of $\vF^{(L)}_{i,t}$.

The global-averaged scores $\bar{\vF}_{i,t}$ is subsequently fed into the fully-connected (FC) layer activated by the softmax function to deliver the final output $\bm{P}_{i,t + R}$, which is computed by
\begin{align}\label{fc}
\begin{split}
    \mbox{(linear transformation)}\quad \bm{P}_{i, t+R}^\ast & \equiv (P^\ast_{i,t+R,0}, P^\ast_{i,t+R,1})' = \bA \bar{\vF}_{i,t}  + \bb\in\mathcal{R}^{2},
    \\
    \mbox{(activation)}\quad \bm{P}_{i,t + R} & \equiv (P_{i,t+R,0}, P_{i,t+R,1})' = \mathrm{softmax}(\bm{P}_{i, t+R}^\ast)\in\mathcal{R}^{2},
\end{split}
\end{align}
where $R$ is a user-specific forward horizon, $P^\ast_{i,t+R,0}$ and $P^\ast_{i,t+R,1}$ are class scores for the classes of ``negative return'' and ``positive return'', respectively, $\bA \in \mathcal{R}^{2 \times C^{(L)}}$ is the matrix of weight parameters, $\bb \in \mathcal{R}^{2}$ is the vector of bias parameters, and $\mathrm{softmax}(\cdot)$ is the activation function to normalize the class scores $P^\ast_{i,t+R,0}$ and $ P^\ast_{i,t+R,1}$ into the class probabilities $P_{i,t+R,0}$ and $P_{i,t+R,1}$ with $P_{i,t+R,0}>0$, $P_{i,t+R,1}>0$, and $P_{i,t+R,0}+P_{i,t+R,1}=1$.

Finally, we summarize the specification of the ResNet model as follows:
\begin{align}\label{model_resnet}
\bm{P}_{i,t + R}=f(\vF_{i,t,D}; \btheta_{D,R}),
\end{align}
where $\vF_{i,t,D}$ is the input, $\bm{P}_{i,t + R}$ is the output, $f(\cdot;\btheta_{D,R})$ is the function explicitly specified by (\ref{res_1})--(\ref{res}) and (\ref{avg_pool})--(\ref{fc}) altogether, and
$\btheta_{D,R}$ contains all of parameters in (\ref{res_1})--(\ref{res}) and (\ref{fc}).

% For a visual investigation, one can refer to the diagram of ResNet models in Fig ...

For $\bm{P}_{i,t + R}$ in (\ref{model_resnet}), the ResNet ``trader'' takes $P_{i,t+R,1}$ as the probability that the $R$-day return of stock $i$ on day $t+R$ is positive. After assigning the label $y_{i,t+R}=1$ for the image represented by $\vF_{i,t,D}$ if the $R$-day return of stock $i$ on day $t+R$ is positive and $y_{i,t+R}=0$ otherwise, he applies the idea of classification to construct the estimator of $\btheta_{D,R}$ defined as
\begin{align}\label{cross_entropy}
    \begin{split}
        \widehat{\btheta}_{D, R} &= \argmin_{\btheta} \sum_{t\in\mathcal{T}} \sum_{i\in \mathcal{I}} - y_{i,t + R} \log(P_{i,t+R,1}) - (1 - y_{i, t+R}) \log(1 - P_{i,t+R,1})
        \\
        & \equiv \argmin_{\btheta} \sum_{t\in\mathcal{T}} \sum_{i\in \mathcal{I}} \ell(y_{i,t + R}, \vF_{i,t,D}, \btheta),
    \end{split}
\end{align}
where the loss function above is based on the cross-entropy, and $\mathcal{I}$ and $\mathcal{T}$ are the sets of stocks and days for training, respectively, To compute $\widehat{\btheta}_{D, R}$ with massive data, the adaptive moment estimation (Adam) algorithm \citep{Kingma2015AdamAM} is applied; see Algorithm \ref{alg:adam} in the Appendix B for its details.
Using $\widehat{\btheta}_{D,R}$, the ResNet ``trader'' predicts $y_{i,t+R}$ by $\widehat{y}_{i,t+R}$ on a testing day $t$, where
\begin{align}\label{esti_y}
\widehat{y}_{i, t+R}  = \begin{cases}
    1, &\mbox{if } \widehat{P}_{i,t+R,1} > \widehat{P}_{i,t+R,0},\\
    0, &\text{otherwise,}
    \end{cases}
\end{align}
with $\widehat{\bm{P}}_{i,t + R}\equiv (\widehat{P}_{i,t+R,0},\widehat{P}_{i,t+R,1})'=f(\vF_{i,t,D}; \widehat{\btheta}_{D,R})$ being the predicted value of $\bm{P}_{i,t + R}$. The predictor in (\ref{esti_y}) is rational, since the $R$-day return of stock $i$ on day $t+R$ is most likely positive if $\widehat{P}_{i,t+R,1} > \widehat{P}_{i,t+R,0}$ (i.e., $\widehat{P}_{i,t+R,1}>1/2$).

\subsection{The Comparison with the CNN ``Trader''}

The CNN ``trader'' in \cite{jiang2023re} applies the same method as the ResNet ``trader'' to find the general price patterns in $D$-day images, except that the former uses the CNN model instead of ResNet model to capture $\bm{P}_{i,t + R}$. Although the CNN model is able to discover complex and non-linear patterns automatically by the recursive extraction, it is challenging to train the deeper CNN model (with more blocks) due to the notorious problem of vanishing or exploding gradients \citep{bengio1994learning}.  Specifically, regarding the chain rule for derivatives, the gradients used for optimizing parameters of shallow layers (e.g., $\btheta^{(1)}$) vanish or explode exponentially within a deep architecture of CNN model. This problem directly destabilizes the training process of an in-depth CNN model, leading to a poor model performance. Unlike the CNN model, the ResNet model solves the above problem by introducing the key component $\mathbb{R}\big(\vF_{i,t}^{(l - S)} \big)$ in \eqref{res} to reformulate blocks as learning residual functions concerning the previous block outputs, rather than learning unreferenced functions. Due to the incorporation of residual learning, the gradients can flow easily back through the network in the ResNet model, effectively tackling the problem of vanishing or exploding gradients. Consequently, it is easy for the ResNet ``trader'' to use the deep ResNet model, which is more capable of discovering complex useful price patterns in $D$-day images than the shallow CNN model used by the CNN ``trader''.

\begin{figure}[!htp]
    \centering
    \includegraphics[width = 0.7\linewidth]{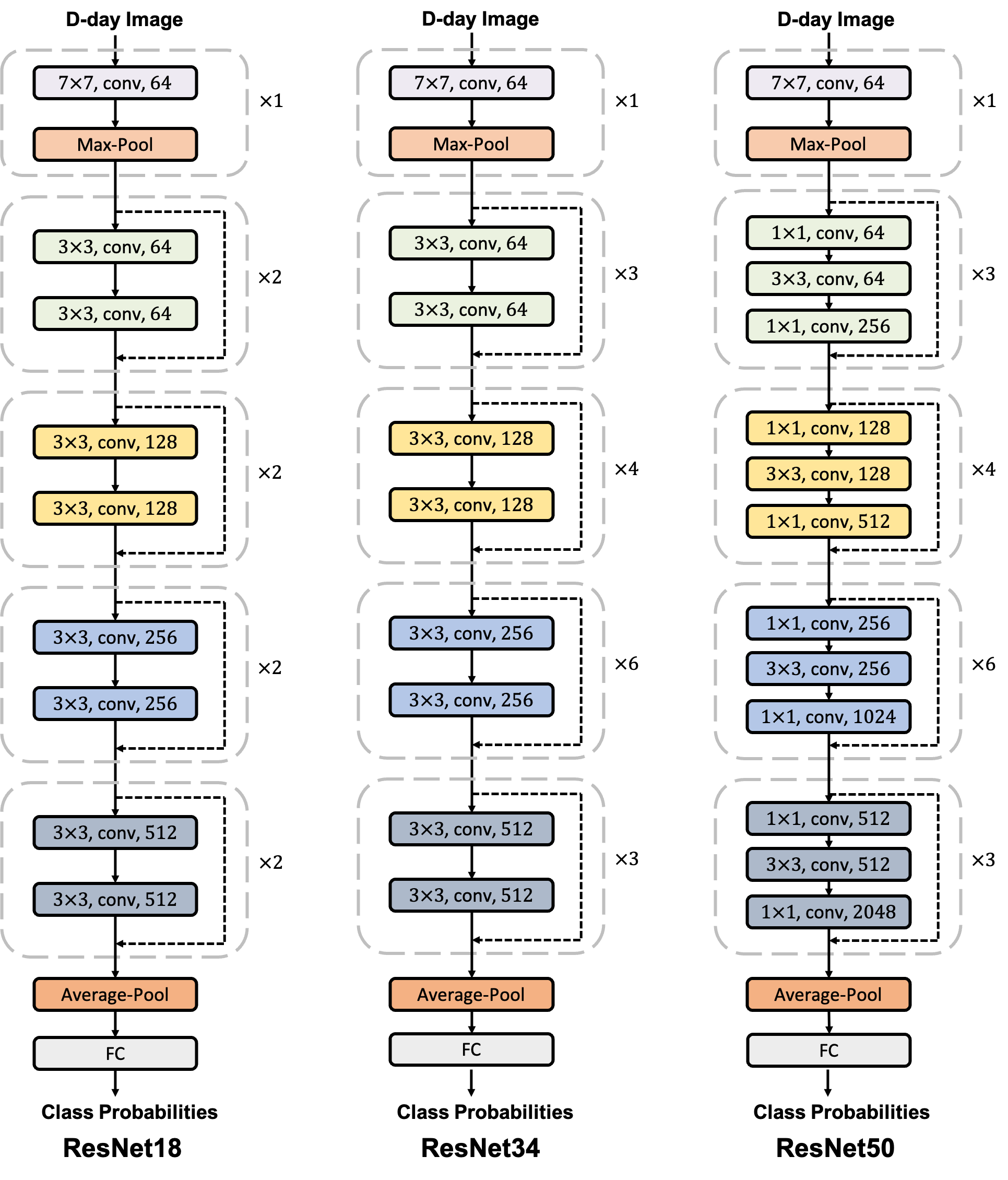}
    \caption{(Color online) Diagrams of three ResNet models: ResNet18 (left), ResNet34 (middle), and ResNet50 (right). These models are built with different settings of convolution (abbreviated as ``conv'' in this figure). The size of the convolution filters in each convolution operation, denoted by the number preceding ``conv'', determines the shape of the output feature map (i.e., $H^{(l)}$ and $W^{(l)}$). The number of convolution filters in each convolution operation, denoted by the number following ``conv'', determines the number of channels of the output feature map (i.e., $C^{(l)}$). The final output through the convolution block is reduced into a vector by average pooling and then fed into a FC layer, which generates the class probabilities of ``negative return'' and ``positive return.'' Note that all activation operations are ignored in this figure, and conventionally, the number of layers in each model only counts the convolution and FC layers, as the max-pooling and average-pooling layers include no trainable parameters.}
    \label{fig:network}
\end{figure}

In line with the recommendation of \cite{he2016resnet}, the ResNet ``trader'' uses 18-, 34-, and 50-layer ResNet models (denoted as ResNet18, ResNet34, and ResNet50), which have approximately 11 million, 21 million, and 23 million parameters, respectively. See Fig\,\ref{fig:network} for the diagrams of these three models. In contrast, the deepest CNN model used by the CNN ``trader'' in \cite{jiang2023re} comprises only 5 layers with less than 3 million parameters.

\subsection{Regularization Implementations}\label{sec:implement}

In this section, we give the details of regularization for the ResNet model. For more insight discussions on regularization for neural networks,
one can refer to \cite{Gu2020EmpiricalAP} and \cite{yang2024asset}.

To alleviate the common issue of overfitting in neural networks, we adopt the standard approach of partitioning the $D$-day image data set into three disjoint parts: training sample, validation sample, and testing sample. The training and validation samples are employed to obtain $\widehat{\btheta}_{D, R}$. Specifically, we compute $\btheta^{(l)}$ at the $l$-th iteration in Algorithm \ref{alg:adam} using the training sample, with the initial learning rate of $10^{-5}$ and batch size of 128 as in \cite{jiang2023re}. Then, we calculate the corresponding validation sample error, which is the value of loss function in \eqref{cross_entropy} with $\btheta=\btheta^{(l)}$ based on the validation sample.
Under the early stopping principle to prevent overfitting, we halt the above iteration process,  when the validation sample error starts to increase over several iterations. Among all of $\btheta^{(l)}$ before stopping, we take the one with the smallest validation sample error to be the estimator $\widehat{\btheta}_{D, R}$. Once $\widehat{\btheta}_{D, R}$ is obtained to deliver a trained ResNet model, the testing sample is used to evaluate the true out-of-sample prediction performance of our trading strategies.

% \subsubsection{Ensemble Learning}

In addition to the risk of overfitting, another weakness caused by the rich parameterization of neural networks is that the estimators of network parameters often get trapped in local optima rather than global optima, due to the use of suboptimal initial values of network parameters generated
from a random seed. To alleviate the randomness caused by initialization, we adopt an ensemble approach by considering
multiple random seeds (say, e.g., $5$), each of which leads to a trained ResNet model to implement the out-of-sample prediction.
Then, the ensemble approach averages the out-of-sample prediction results from all of trained ResNet models.

\section{Enhancement of Price Trend Strategies}\label{sec:TWMA}

Due to the rich parameters and high level of abstraction of ResNet,  the ResNet ``trader'' can automatically detect the general price patterns from $D$-day images. From a trading perspective, interpreting these predictive patterns is important. In this section, we provide
a way to visually see these general price patterns, from which we further propose a new method to enhance the trading signals of the existing strategies.

\subsection{The Localization Maps}

%Due to the rich parameters and high level of abstraction of ResNet,  the ResNet ``trader'' can automatically detect the general price patterns from D-day images. However,

In terms of interpretability, the first natural but challenging question is: What are the general price patterns detected by the ResNet model? This question is unavoidable, if the traders in real world aim to make use of these general price patterns.
In this section, we tackle the above question by constructing two localization maps, based on the smooth gradient-weighted class activation mapping (smooth Grad-CAM) method in \cite{omeiza2019smooth}.

%zhou2016learning,  wang2020score

The smooth Grad-CAM method stems from the gradient-weighted class activation mapping (Grad-CAM) method (\citealp{selvaraju2017grad}). In line with the downstream architecture of ResNet from the final feature map $\vF^{(L)}_{i,t}$ to class probabilities
$P_{i, t+R, 0}$ and $P_{i, t+R, 1}$, the Grad-CAM method assumes the following linear approximation:
\begin{align}\label{linear_approx}
    P_{i, t+R, j} \approx \sum_{c = 1}^{C^{(L)}} G_{i,t,j}(c) \sum_{h = 1}^{H^{(L)}} \sum_{w = 1}^{W^{(L)}} F_{i,t}^{(L)} (h, w, c)
\end{align}
for $j=0$ and 1, where $F_{i,t}^{(L)} (h, w, c)$ is the $(h,w,c)$-th entry of $\vF^{(L)}_{i,t}$, and $G_{i,t,j}(c)$  represents the importance level of $c$-th channel of $\vF_{i,t}^{(L)}$ to the class probability $P_{i, t+R, j}$. Clearly, under the approximation in (\ref{linear_approx}),
the contribution of $F_{i,t}^{(L)} (h, w, c)$ to $P_{i, t+R, j}$ is invariant across $h$ and $w$ on a given $c$-th channel\footnote{This spatially invariant property is linked to the average-pooling operation applied in \eqref{avg_pool}. The average-pooling operation, which reduces the final feature map into channel-wise scores, has been proven effective in the field of computer vision; see \cite{ConvNet} and references therein. Consequently, this effectiveness lends rationality to the design of the linear approximation in \eqref{linear_approx}.}. Re-writing (\ref{linear_approx}) as
\begin{align*}
    P_{i, t+R, j} \approx \sum_{h = 1}^{H^{(L)}} \sum_{w = 1}^{W^{(L)}} \Big[\sum_{c = 1}^{C^{(L)}} G_{i,t,j}(c) F_{i,t}^{(L)} (h, w, c)\Big],
\end{align*}
it is evident that the value of $\sum_{c} G_{i,t,j}(c) F_{i,t}^{(L)} (h, w, c)$ represents the importance level of extracted feature at $(h,w)$-th spatial location for explaining $P_{i, t+R, j}$, leaving $\{G_{i,t,j}(c): c = 1, \dots, C^{(L)}\}$ as the only undetermined terms.

%\cite{InceptionNet}, \cite{he2016resnet}, \cite{DenseNet},

A simple but effective choice for $G_{i,t,j}(c)$ is the global weighted average gradient \citep{Aditya2018camplus} defined as
\begin{align}\label{global_gradient}
    G_{i,t,j}(c) = \sum_{h = 1}^{H^{(L)}} \sum_{w = 1}^{W^{(L)}} \alpha_{i,t,j}(h,w,c) \text{ReLU}(g_{i,t,j} (h, w, c)),
\end{align}
where the gradient $g_{i,t,j}(h,w,c) = \partial P_{i,t+R,j}/\partial F^{(L)}_{i,t} (h,w,c)$ is the first-order derivative of $P_{i,t+R,j}$ with respect to $F^{(L)}_{i,t} (h,w,c)$, and $\alpha_{i,t,j}(h,w,c)$ is the spatial weight used for normalizing those gradients in (\ref{global_gradient}). Note that $g_{i,t,j}(h,w,c)$ represents the sensitivity of the class probability $P_{i,t+R,j}$ to small changes in $F^{(L)}_{i,t} (h,w,c)$.
Hence, a larger value of $|g_{i,t,j}(h,w,c)|$ indicates a larger impact of $F^{(L)}_{i,t}(h,w,c)$ on $P_{i,t+R,j}$, so the extracted feature at $(h,w,c)$-th location is more important for explaining $P_{i,t+R,j}$. This argument should have motivated us to replace $\text{ReLU}(g_{i,t,j} (h, w, c))$ with $|g_{i,t,j}(h,w,c)|$ to define $G_{i,t,j}(c)$ in (\ref{global_gradient}). However, since $g_{i,t,0}(h,w,c)=-g_{i,t,1}(h,w,c)$ due to the fact that $P_{i,t+R,0}+P_{i,t+R,1}=1$, we have $|g_{i,t,0}(h,w,c)|=|g_{i,t,1}(h,w,c)|$. Therefore, the use of $|g_{i,t,j}(h,w,c)|$ is problematic resulting from an invariant gradient impact on both $P_{i,t+R,0}$ and $P_{i,t+R,1}$. To circumvent this issue, we take $\text{ReLU}(g_{i,t,j} (h, w, c))$ rather than $|g_{i,t,j}(h,w,c)|$ to define $G_{i,t,j}(c)$ in (\ref{global_gradient}). Owing to the presence of $\text{ReLU}(\cdot)$, only the positive $g_{i,t,0} (h, w, c)$ is applied to reflect the importance level of extracted feature at $(h,w,c)$-th location for explaining $P_{i,t+R,0}$, whereas the negative $g_{i,t,0} (h, w, c)$ (i.e., the positive $g_{i,t,1} (h, w, c)$) has the same mission but to serve for $P_{i,t+R,1}$. For the spatial weight $\alpha_{i,t,j}(h,w,c)$ in (\ref{global_gradient}), we follow \cite{Aditya2018camplus} to set
\begin{align}\label{spatial_weight}
    \alpha_{i,t,j}(h,w,c) & = \frac{g^{(2)}_{i,t,j}(h,w,c)}{2 g^{(2)}_{i,t,j}(h,w,c) + \sum_{h = 1}^{H^{(L)}} \sum_{w = 1}^{W^{(L)}} F^{(L)}_{i,t} (h,w,c) g^{(3)}_{i,t,j}(h,w,c)},
\end{align}
where $g^{(2)}_{i,t,j}(h,w,c)$ and $g^{(3)}_{i,t,j}(h,w,c)$ are the second- and third-order derivatives of $P_{i,t+R,j}$ with respect to $F^{(L)}_{i,t} (h,w,c)$, respectively. The formula in \eqref{spatial_weight} is derived by substituting \eqref{global_gradient} into \eqref{linear_approx} and taking partial derivative of both sides with respect to $F^{(L)}_{i,t} (h,w,c)$ twice; see \cite{Aditya2018camplus} for more details.

Although the Grad-CAM method opens up the ``black-box'' mechanism of class probabilities, it has an instability issue on $G_{i,t,j}(c)$.
This is because due to the highly nonlinearity of ResNet, the partial derivatives used in the Grad-CAM method tend to fluctuate sharply at small scales, resulting in unstable $G_{i,t,j}(c)$ \citep{smilkov2017smoothgrad}. To solve this issue, \cite{omeiza2019smooth} propose a smooth Grad-CAM method. This method first generates $B$ noised images $\{\vF_{i,t,D,b}: b = 1, \dots, B\}$ by adding Gaussian noise to the original image $\vF_{i,t,D}$. Then, for each noised image $\vF_{i,t,D,b}$, it computes the corresponding $\vF^{(L)}_{i,t,b}$ and class probabilities $P_{i,t+R,0,b}$ and $P_{i,t+R,1,b}$, leading to the smoothed global weighted average gradient:
\begin{align}\label{smooth_global_gradient}
    \widetilde{G}_{i,t,j}(c) = \sum_{h = 1}^{H^{(L)}} \sum_{w = 1}^{W^{(L)}} \widetilde{\alpha}_{i,t,j}(h,w,c) \text{ReLU}( \widetilde{g}_{i,t,j} (h, w, c)),
\end{align}
where
\begin{align*}
    \widetilde{g}_{i,t,j} (h, w, c) & = \frac{1}{B} \sum_{b=1}^{B} g_{i,t,j,b} (h, w, c),
    \\
    \widetilde{\alpha}_{i,t,j}(h,w,c) & = \frac{\frac{1}{B} \sum_{b=1}^{B} g^{(2)}_{i,t,b,j}(h,w,c)}{\frac{2}{B} \sum_{b=1}^{B} g^{(2)}_{i,t,j,b}(h,w,c) + \frac{1}{B} \sum_{b=1}^{B} \sum_{h = 1}^{H^{(L)}} \sum_{w = 1}^{W^{(L)}} F^{(L)}_{i,t,b} (h,w,c) g^{(3)}_{i,t,j,b}(h,w,c)}.
\end{align*}
Here, $g_{i,t,j,b} (h, w, c)$, $g^{(2)}_{i,t,j,b} (h, w, c)$, and $g^{(3)}_{i,t,j,b} (h, w, c)$ are the first-, second-, and third-order derivatives of $P_{i,t+R,j, b}$ with respect to $F^{(L)}_{i,t,b} (h,w,c)$, respectively. Compared with $G_{i,t,j}(c)$ in (\ref{global_gradient}), $\widetilde{G}_{i,t,j}(c)$ relies on the averaged partial derivatives obtained from $B$ different noised images in a neighborhood of  $\vF_{i,t,D}$,
so it is a robust measurement of channel-wise importance.

Next, based on $\{\widetilde{G}_{i,t,j}(c): c = 1, \dots, C^{(L)}\}$, we define the localization matrix $\widetilde{\bW}_{i,t,j}\in\mathcal{R}^{H^{(L)}\times W^{(L)}}$, where the $(h,w)$-th entry of $\widetilde{\bW}_{i,t,j}$ is
\begin{align}\label{importance_matrix}
    \widetilde{W}_{i,t,j}(h,w) = \mathrm{ReLU}\Big[\sum_{c=1}^{C^{(L)}} \widetilde{G}_{i,t,j}(c) F_{i,t}^{(L)}(h,w,c)\Big].
\end{align}
As explained above, the value of $\sum_{c} \widetilde{G}_{i,t,j}(c) F_{i,t}^{(L)} (h, w, c)$ represents the importance level of extracted feature at $(h,w)$-th spatial location for explaining $P_{i, t+R, j}$. To better highlight the important regions in the image, we follow \cite{selvaraju2017grad} to implement
$\mathrm{ReLU}(\cdot)$ in (\ref{importance_matrix}). Finally, we apply the bilinear interpolation method to resize $\widetilde{\bW}_{i,t,0}$ and  $\widetilde{\bW}_{i,t,1}$ into the ``downward'' localization map $\bW_{i,t,0} \in \mathcal{R}^{224 \times 224}$ and ``upward'' localization map $\bW_{i,t,1} \in \mathcal{R}^{224 \times 224}$, respectively.
By construction, each spatial location of $\bW_{i,t,0}$ (or $\bW_{i,t,1}$) is assigned with a value of importance, which represents how important is this location for learning $P_{i,t+R,0}$ (or $P_{i,t+R,1}$). Hence, by
overlaying $\bW_{i,t,0}$ (or $\bW_{i,t,1}$) on $D$-day image $\vF_{i,t,D}$, we can locate the important regions of $\vF_{i,t,D}$ to exhibit the general price patterns detected by the ResNet ``trader''. In practice, once the ResNet model is well-trained, we can obtain $\bW_{i,t,0}$ and $\bW_{i,t,1}$
by replacing $\btheta_{D,R}$ with $\widehat{\btheta}_{D, R}$ in \eqref{cross_entropy}.

For a visual illustration, Figure \ref{fig:CAM_kline} presents some localization maps (from our empirical analysis in Section \ref{sec:empirical} below) by overlaying them on the related $D$-day images. From this figure, we have some common observations in both ``downward'' and ``upward'' localization maps. First, two localization maps for the 5- and 20-day images demonstrate the crucial role of the moving average price line in prediction of the direction of stock price one day ahead (i.e., $R=1$), since the crosses of OHLC bars and moving average price line in these images are highlighted. Second, two localization maps for the 60-day image concentrate mainly on the right-hand side of the image, which contains the most recent price and volume data.
Along with the above common observations, we also find some significant differences in two localization maps regardless of the choice of $D$-day image. Specifically, the ``upward'' localization map consistently emphasizes the impact from daily trading volume bars, whereas the ``downward'' localization map focuses solely on the price-related objects, such as OHLC bars and the moving average price line. Overall, the findings from Figure \ref{fig:CAM_kline} indicate that the general price patterns detected by the ResNet ``trader'' vary with the choice of $D$-day image and the target of probability. As expected, these general price patterns may not closely align with a pre-specific price pattern or trading rule, but they should carry some useful information that is not readily available or visible to the traders in real world.

\begin{figure}[!h]
    \centering
    \includegraphics[width = 0.95\linewidth]{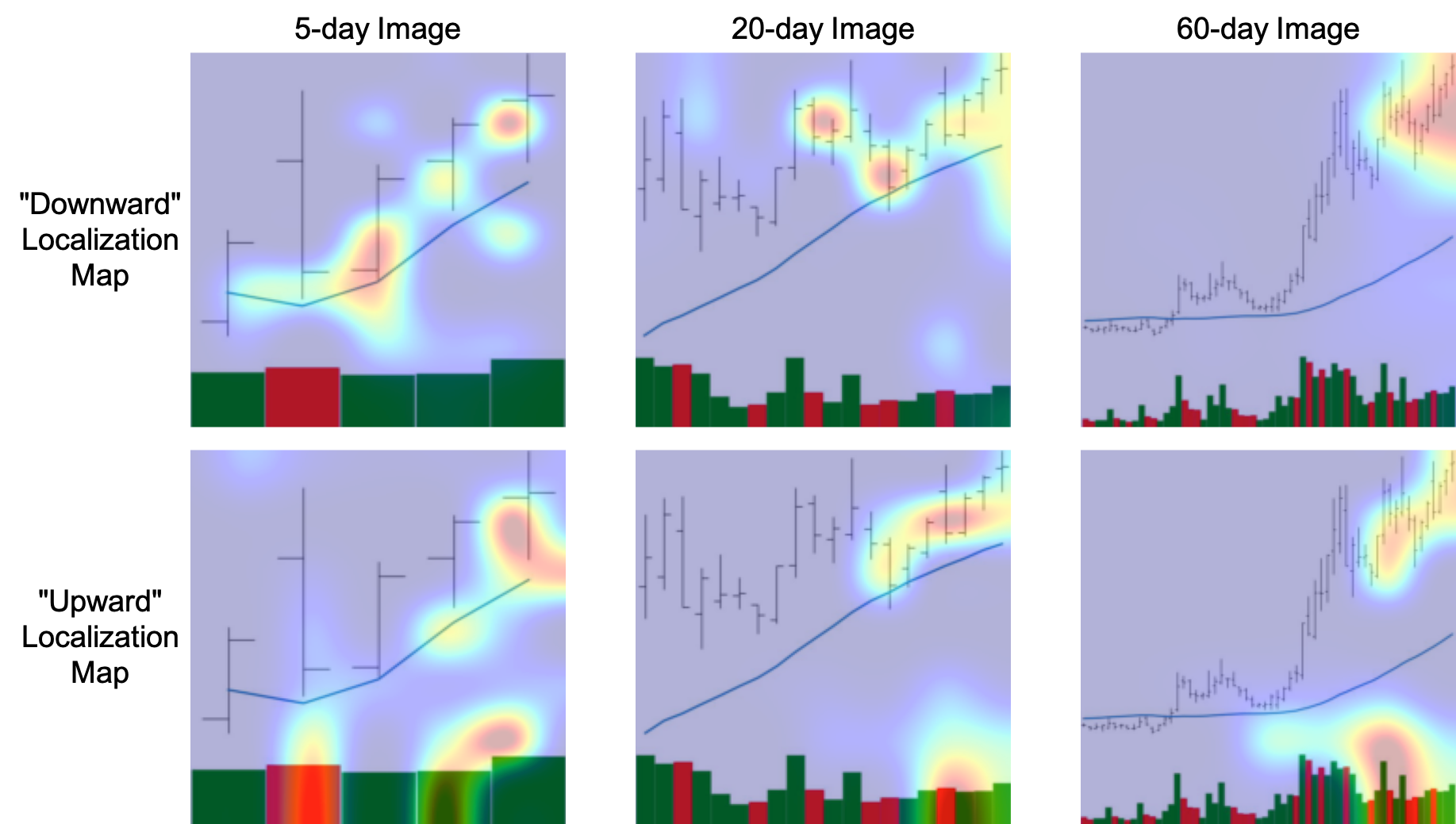}
    \caption{(Color online) The ``downward'' localization map $\bW_{i,t,0}$ and ``upward'' localization map $\bW_{i,t,1}$ generated by the ResNet34 model, where the stock $i$ is Ping An Bank, the day $t$ is January 5, 2015, and the ResNet34 model with $R=1$ is used to predict the direction of stock price one day ahead.
    Here, the 5-, 20-, 60-day images come from the daily market data of Ping An Bank, with January 5, 2015 as the ending day.
    The brighter regions in ``downward'' (or ``upward'') location maps correspond to higher values in $\bW_{i,t,0}$ (or $\bW_{i,t,1}$), and the gray areas indicate zero values in $\bW_{i,t,0}$ (or $\bW_{i,t,1}$).}
    \label{fig:CAM_kline}
\end{figure}

\subsection{The Triple-I Weighted Moving Average Method}

After visualizing the general price patterns detected by the ResNet ``trader'', the traders face a follow-up attractive question: How can we make use of these detected general price patterns to enhance our own trading strategies? Finding a desirable answer to this question is crucial, since it not only provides a strong evidence to support the viewpoint that ResNet ``trader'' is capable of extracting valuable information from price chart images, but also shows how the traders can benefit from the tools of artificial intelligence in the era of big data. In this section, we tackle the aforementioned question via a novel weighted moving average method.

Our weighted moving average method first aims to construct a vector of image-induced importance (III, or so-called triple-I) weights from the localization matrix $\widetilde{\bW}_{i,t,j}$ through a three-step procedure. In step one, we compress $\widetilde{\bW}_{i,t,j}$ vertically into a $W^{(L)}$-dimensional weighting vector $\bomega_{i,t,j}^{\ast}$, where the $w$-th entry of $\bomega_{i,t,j}^{\ast}$ is
\begin{align*}
    \bomega_{i,t,j}^{\ast}(w) = \frac{1}{H^{(L)}} \sum_{h = 1}^{H^{(L)}} \widetilde{W}_{i,t,j}(h,w),
\end{align*}
which is the average of all entries in $w$-th column of $\widetilde{\bW}_{i,t,j}$. In step two, we apply the linear interpolation method to resize $\bomega_{i,t,j}^{\ast}$ into a $D$-dimensional weighting vector $\bomega^{\ast\ast}_{i,t,j} \equiv (\omega_{i, t, j, D}^{\ast\ast}, \dots, \omega_{i,t, j, 1}^{\ast\ast})'$.
%\begin{align*}
%    \bomega^{\ast\ast}_{i,t,j} \equiv (\omega_{i, t, j, D}^{\ast\ast}, \dots, \omega_{i,t, j, 1}^{\ast\ast})'.
%\end{align*}
In step three, we normalize $\bomega^{\ast\ast}_{i,t,j}$ to propose the triple-I weighting vector
\begin{align*}
    \bomega_{i,t,j} \equiv (\omega_{i,t,j, D}, \dots, \omega_{i,t,j,1})' = \frac{1}{\sum_{d = 1}^D \omega_{i,t,j, d}^{\ast\ast}} \bomega^{\ast\ast}_{i,t,j},
\end{align*}
where the entries $\omega_{i,t,j,d}$, $d=1,\dots, D$, are called triple-I weights. The logic of each aforementioned step is rational. The first compression step computes the overall level of importance at each temporal location of $\widetilde{\bW}_{i,t,j}$ for explaining the class probability $P_{i,t+R,j}$. The second resizing step aligns the temporal dimension of weighting vector with that of the $D$-day image. The third normalization step ensures that the sum of all triple-I weights is one.

By construction, the value of $\omega_{i,t,j,d}$ reflects the importance of
day $t-d+1$ for explaining the class probability $P_{i,t+R,j}$, where $P_{i,t+R,0}$ (or $P_{i,t+R,1}$) is the probability that $y_{i, t+R} = 0$ (or 1), with $y_{i, t+R}$ defined as in (\ref{cross_entropy}). Based on this, our weighted moving average method adopts $\bomega_{i,t,0}$ and $\bomega_{i,t,1}$ to construct the weighted moving average trading signal:
\begin{align}\label{IWA_signal}
    \overline{x}_{i,t} =
    \begin{cases}
        \sum_{d = 1}^D \omega_{i, t, 0, d} x_{i, t - d + 1}, & \text{if}\,\,\, y_{i, t+R} = 0, \\
        \sum_{d = 1}^D \omega_{i, t, 1, d} x_{i, t - d + 1}, & \text{if}\,\,\, y_{i, t+R} = 1,
    \end{cases}
\end{align}
where $x_{i, t}$ is the original trading signal used by the traders for stock $i$ on day $t$ to form their own trading strategies. In (\ref{IWA_signal}), we select $\bomega_{i,t,0}$ (or $\bomega_{i,t,1}$) to construct $\overline{x}_{i,t}$, depending on whether $y_{i, t+R} = 0$ (or 1). This selection aligns with the fact that $\bomega_{i,t,0}$ (or $\bomega_{i,t,1}$) naturally captures the importance of all temporal locations from $t-D+1$ to $t$ in predicting the probability that $y_{i, t+R} = 0$ (or 1). Since the construction of $\overline{x}_{i,t}$ relies on the triple-I weights, our method above is termed the triple-I weighted moving average (TWMA) method. This TWMA method is new to the literature. It produces the TWMA-based trading signal $\overline{x}_{i,t}$, which combines the knowledge of both machine and human via the use of the triple-I weights (guided by machine) to weighted moving average original trading signal (built up from human domain knowledge). In practice, we can easily compute $\overline{x}_{i,t}$, when $\widehat{\btheta}_{D, R}$ in \eqref{cross_entropy} and $\widehat{y}_{i, t+R}$ in \eqref{esti_y} from a well-trained ResNet model are taken to replace $\btheta_{D,R}$ and $y_{i, t+R}$, respectively. Below, our empirical task is to investigate whether the trading strategies based on $\overline{x}_{i,t}$ outperform those based on $x_{i,t}$. This empirical investigation is key to determine whether the TWMA method takes a first step to answer our targeted question.

\section{Empirical Performance of TWMA-based Strategies}\label{sec:empirical}

In this section, we apply our TWMA method to propose new price trend trading strategies in the Chinese stock market. Our goal is to examine whether these new TWMA-based strategies perform better than their existing counterparts used by the traders in real world.

\subsection{Data}

We consider daily stock data from the Wind database (\url{https://www.wind.com.cn}) for A-share stocks listed on the Shanghai and Shenzhen Stock Exchanges in China. To avoid anomalies associated with small-cap stocks, we primarily focus on the largest $1,800$ A-share stocks by market capitalization. Our data collection period spans from January 1, 2014, to May 1, 2023, giving a dataset comprising over 2 million $D$-day images for each choice of $D$.
%\footnote{A relaxation of the size filter will be investigated in Sec \ref{sec:stock_size}.}.

Moreover, we use the same data pre-processing techniques as in \cite{jiang2023re}. First, we normalize the first day closing price to one and calculate each subsequent daily closing price using the corresponding daily return in each image. This scale proportion to the closing price is applied to the related opening, high, and low prices. Second, we normalize all images by using the mean and standard deviation of pixel values from the images in the training sample. Third, we randomly select training and validation samples and do not chronologically re-train the ResNet model. To be specific, we train and validate ResNet model only once using the data from 2014 to 2020, in which $70\%$ of the images are randomly selected for the training sample and the remaining $30\%$ for the validation sample. The data from January 1, 2021 to May 1, 2023 are selected as the testing sample, which is used for truly out-of-sample evaluation.

\subsection{Comparison Methods}

To evaluate the effectiveness of the TWMA method in enhancing technical trading signals, we consider five well-known original technical trading signals: 2-12 momentum (MOM), 1-month short-term reversal (STR), 1-week short-term reversal (WSTR), trend (TREND, \citealp{han2016trend}), and formulaic alphas (ALPHA, \citealp{kakushadze2016alpha}).

The first four trading signals are analyzed in \cite{jiang2023re}. For the ALPHA signal, we generate it by utilizing 101 price-and-volume-based formulaic alphas in \cite{kakushadze2016alpha}. Due to the highly skewed and leptokurtic alphas, we first use the rank normalization method in \cite{Gu2020EmpiricalAP} to normalize the alphas into the interval $(-1, 1)$. Then, we aggregate those alphas via a cross-sectional regression on day $t$:
\begin{align*} %\label{cross_reg}
    r_{i,t+R} = \beta_{0,t} + \sum_{k = 1}^{101} \beta_{k,t} A_{i,t,k} + \epsilon_{i,t} \,\,\, \text{for}\,\,\, i = 1, \dots, N,
\end{align*}
where $N=1800$, $r_{i,t+R}$ is the $R$-day return of stock $i$ on day $t+R$, $A_{i,t,k}$ is the $k$-th rank-normalized alpha, $\epsilon_{i,t}$ is the error term with zero mean, and $\beta_{0,t}$ and $\beta_{k,t}$ are intercept and regression coefficients, respectively. To avoid the potential high-dimensional issue in the regression, we use the adaptive LASSO method \citep{Zou2006Adaptive} to estimate coefficients $\{\beta_{k,t}: k = 0, \dots, 101\}$ for each $t$. Finally, the ALPHA signal aggregated from 101 alphas on day $t$ is constructed by:
\begin{align*} %\label{cross_pred}
    \widehat{r}_{i,t+R} = \widetilde{\beta}_{0,t} + \sum_{k = 1}^{101} \widetilde{\beta}_{k,t} A_{i,t,k} \,\,\, \text{with}\,\,\, \widetilde{\beta}_{k,t} = \frac{1}{252} \sum_{m = 1}^{252} \widehat{\beta}_{k,t-m},
\end{align*}
where $\widehat{\beta}_{k,t}$ is the adaptive LASSO estimator of $\beta_{k,t}$, and $\widetilde{\beta}_{k,t}$ is its one-year (roughly 252 trading days) moving average. Notably, our way to construct the ALPHA signal is similar to that in \cite{han2016trend}, which aggregates different normalized moving average prices to generate the TREND signal.

For each signal above, we apply the TWMA method to obtain its TWMA-based counterpart. To be specific,
we directly get the TWMA-based MOM, STR, and WSTR signals as in \eqref{IWA_signal}. However, this direct approach is unsuitable for TREND and ALPHA
signals, which necessitate cross-sectional regressions to aggregate diverse stock-level information. Therefore,
we deal with these two signals by replacing the original regressors in the regression with their TWMA-based counterparts.
For instance, when applying the TWMA method to the ALPHA signal, we define the cross-sectional regression as follows:
\begin{align*} %\label{iwa_cross_reg}
    r_{i,t+R} = \beta_{0,t}^\ast + \sum_{k = 1}^{101} \beta_{k,t}^\ast A_{i,t,k}^\ast + \epsilon_{i,t}^\ast \,\,\, \text{for}\,\,\, i = 1, \dots, N,
\end{align*}
where $A_{i,t,k}^\ast$ is the TWMA-based $A_{i,t,k}$ as formed in \eqref{IWA_signal}, $\epsilon_{i,t}^\ast$ is the error term with zero mean, and $\beta_{0,t}^\ast$ and $\beta_{k,t}^\ast$ are intercept and regression coefficients, respectively. With $\{\widehat{\beta}_{k,t}^\ast: k = 0, \dots, 101\}$ estimated by the adaptive LASSO method, the TWMA-based ALPHA signal on day $t$ is then given by
\begin{align*} %\label{iwa_cross_pred}
    \overline{r}_{i,t+R}^\ast = \widetilde{\beta}_{0,t}^\ast + \sum_{k = 1}^{101} \widetilde{\beta}_{k,t}^\ast A_{i,t,k}^\ast \,\,\, \text{with}\,\,\, \widetilde{\beta}_{k,t}^\ast = \frac{1}{252} \sum_{m = 1}^{252} \widehat{\beta}_{k,t-m}^\ast.
\end{align*}
Analogously, we can use the above approach to obtain the TWMA-based TREND signal.

Notably, the CNN model (or any other artificial intelligence model) can replace the ResNet model to propose the triple-I weights for proceeding the TWMA-based trading signal. For ease of presentation, the TWMA method based on the ResNet (or CNN) model is
termed the TWMA-RN (or TWMA-CN) method. By conducting a comparison between the TWMA-RN and TWMA-CN methods, we can ascertain whether the ResNet ``trader'' is smarter than the CNN ``trader'' in \cite{jiang2023re} for extracting information from price chart images. Moreover, we also consider the exponentially weighted moving average (EWMA) method as another competitor for the TWMA-RN method. The EWMA method replaces both of triple-I weights
$\omega_{i, t, 0, d}$ and $\omega_{i, t, 1, d}$ in \eqref{IWA_signal} with the exponentially decayed weights $\omega_{d}^{\dag}$ for all $i$ and $t$ to generate new EMWA-based trading signals, where
$$\omega_{d}^{\dag}=\frac{\lambda(1-\lambda)^{d-1}}{\sum_{d=1}^{D} \lambda(1-\lambda)^{d-1}}$$
for $d=1,\dots, D$, and $\lambda\in(0, 1)$ is the smoothing parameter. The comparison between the TWMA-RN and EWMA methods aims to determine whether the image-induced weights determined by the ResNet model are more informative than the non-data-driven exponentially decayed weights.

Finally, we construct portfolios using each aforementioned trading signal. Specifically, we sort all stocks into ten deciles based on the value of each signal on day $t$, and then we construct the long-short portfolio by buying the 10\% highest ranking stocks (decile 10) and selling the 10\% lowest ranking stocks (decile 1). Using equal weights, we reconstitute portfolios every $R$ days (that is, the holding-period for each portfolio coincides with the forecast horizon), as done in \cite{jiang2023re}. In summary, we have the following portfolios from all of considered methods:

(i) The MOM, STR, WSTR, TREND, and ALPHA portfolios from five original trading signals;

(ii) The $\mbox{RN}x$-$\mbox{D}y/\mbox{R}z$ portfolios from five TWMA-RN-based trading signals, where the notation ``$\mbox{RN}x$-$\mbox{D}y/\mbox{R}z$'' indicates that the ResNet$x$ model uses $y$-day images to predict subsequent $z$-day holding period returns;

%which are generated via a ResNet$x$ model to analyze $y$-day images for predicting subsequent $z$-day holding period returns;

(iii) The $\mbox{CN}$-$\mbox{D}y/\mbox{R}z$ portfolios from five TWMA-CN-based trading signals, where the notation ``$\mbox{CN}$-$\mbox{D}y/\mbox{R}z$'' indicates that the CNN model uses $y$-day images to predict subsequent $z$-day holding period returns;

(iv) The EWMA$m$ portfolios from five EWMA-based trading signals, where the EWMA method uses the smoothing parameter $\lambda=m/100$.

Below, we take $x=18$, 24, and $50$ with respect to three ResNet models used in \cite{he2016resnet}. Following \cite{jiang2023re}, we set $y=5$, 20, and $60$ for the image to account for the weekly, monthly, or quarterly price and volume information, respectively. Since the image-based return predictions and technical analysis signals are likely effective over relatively short horizons (\citealp{jiang2023re}), we focus our analysis on the performance of one-day and one-week (i.e., $z=1$ and 5) holding periods. As suggested by the document of ``RiskMetrics'', we choose $m=94$ (i.e., $\lambda=0.94$). In addition, we also consider $m=98$ and $90$ to see how sensitive is the EWMA method to the choice of $\lambda$.

\subsection{Portfolio Performance}

We evaluate all considered portfolios by analyzing the values of annualized excess return, annualized Sharpe ratio, and monthly turnover (\citealp{Gu2020EmpiricalAP}) on the testing sample, where excess return and Sharpe ratio are computed using the Treasury bill rate as a proxy for the risk-free return. Commonly, Sharpe ratio is the most essential out-of-sample metric for portfolio evaluation since it balances return and risk, while excess return and turnover offer additional perspectives on portfolio performance.

\subsubsection{Impact of Method}\label{sec:iwa_model}

We start our out-of-sample evaluation for portfolios with one-day holding period, which target one-day-ahead returns. Table \ref{tab:daily_TWMA} assesses the predictive performance of all portfolios proposed by different methods. From this table, we can have the following findings:

(i) In each trading signal class, the RN34-D5/R1 method performs the best with the highest value of Sharpe ratio. Specifically, in the classes of MOM, STR, and WSTR, the RN34-D5/R1 method demonstrates at least 219\% higher values of Sharpe ratio and 27\% lower values of turnover than the original method; in the classes of TREND and ALPHA, the RN34-D5/R1 method remains a clear advantage over the original method with 27\% and 73\% higher values of Sharpe ratio, while 33\% and 21\% lower values of turnover, respectively. Moreover, evidenced by higher values of Sharpe ratio and lower values of turnover, both RN34-D20/R1 and RN34-D60/R1 methods always lead to better portfolios than the original method, except for the case of D60 in the class of TREND.

(ii) Each RN50 method exhibits a comparable or slightly worse portfolio performance than its corresponding RN34 method. To be specific, the three RN50 methods have similar values of Sharpe ratio as the three RN34 methods in the classes of MOM and STR, while this comparable phenomenon disappears in the classes of WSTR, TREND, and ALPHA. In terms of maximized value of Sharpe ratio, the three RN18 methods always outperform the original method (except for the TREND class), though they generally perform worse than the corresponding RN34 or RN50 methods.

(iii) Compared with the original method, all (or most) of the CNN and EWMA methods have lower (or higher) values of Sharpe ratio in the classes of TREND and ALPHA (or MOM, STR, and WSTR). While all of the CNN and EWMA methods exhibit comparable performance with the three RN18 methods, their predictive strength is considerably inferior to that of the three RN34 or RN50 methods, regardless of the signal class.

Next, we also evaluate the predictive performance of all methods in an one-week horizon context. Table \ref{tab:weekly_TWMA} presents the results of all portfolios proposed by different methods, where these portfolios with one-week holding period target five-day-ahead returns.
From Table \ref{tab:weekly_TWMA}, we find that the RN34-D20/R5 method performs the best, with 71\%, 65\%, 25\%, and 11\% higher values of Sharpe ratio than the original method in the classes of MOM, STR, WSTR, and TREND, respectively; meanwhile, the RN34-D20/R5 method has a subtly lower value of Sharpe ratio than the original method in the class of ALPHA, but it outperforms the original method by a wide margin regarding the maximized (or minimized) values of excess return (or turnover).
However, the other two RN34 methods do not inherit the advantage of RN34-D20/R5 method, since they have lower values of Sharpe ratio than the original method in most cases. Moreover, unlike the findings from Table \ref{tab:daily_TWMA}, we find from Table \ref{tab:weekly_TWMA} that all of the RN18 and CNN methods do not perform better than the original method in terms of the maximized value of Sharpe ratio; when using D5 and D20 images, the RN50 method outperforms the original method in six out of ten cases; and the EWMA method has a higher value of Sharpe ratio than the original method only in the class of ALPHA. Another different observation from Table \ref{tab:daily_TWMA} is that the EWMA90 method performs the best in the class of ALPHA. Compared with the RN-D20/R5 method, the EWMA90 method earns 15\% higher value of Sharpe ratio but at the expense of 21\% lower value of excess return and 32\% higher value of turnover. Interestingly, both Tables \ref{tab:daily_TWMA} and \ref{tab:weekly_TWMA} exhibit a common observation that regardless of the choice of importance weights, all of weighting methods generally have lower value of turnover than the original method, although this merit sometimes is accompanied by a decline in excess return.

Overall, the aforementioned findings from Tables \ref{tab:daily_TWMA} and \ref{tab:weekly_TWMA} show that the RN34 method based on the use of 5-day image (or $20$-day image) dominates all of other competing methods for one-day-ahead (or five-day-ahead) return predictions. The better performance of RN34 method than the original method is practically crucial, since it indicates that the triple-I weights (through the TWMA approach) can efficiently utilize the general price patterns detected by the ResNet34 model to achieve the prediction enhancement of existing price trend strategies. Meanwhile, the substantial advantage of RN34 (but not RN18) method over the CNN method illustrates the benefits of increasing network depth and parameter size to an appreciable level for identifying essential price patterns, which strengthen the effect of triple-I weights.
However, compared with the RN34 method, the worse performance of RN50 method also reveals that a deeper network could increase the risk of over-fitting, making the TWMA approach less effective. Our analysis demonstrates that the ResNet34 model could well balance the over-fitting risk and prediction accuracy.
In addition, the results from all of RN methods show that the choice of $D$-day image should depend on the forecast horizon.
Typically, we recommend to use 5-day images only for the one-day or other shorter forecast horizons, while we suggest avoiding the use of 60-day images for the one-week or even longer forecast horizons.

\begin{table}[!h]
  \centering
  \caption{One-day long-short portfolio performance across various methods.}
  \scalebox{0.8}{
    \begin{tabular}{cccccccccccccccccccc}
    \toprule
          & \multicolumn{3}{c}{MOM} &       & \multicolumn{3}{c}{STR} &       & \multicolumn{3}{c}{WSTR} &       & \multicolumn{3}{c}{TREND} &       & \multicolumn{3}{c}{ALPHA} \\
\cmidrule{2-4}\cmidrule{6-8}\cmidrule{10-12}\cmidrule{14-16}\cmidrule{18-20}    Method & Ret   & SR    & TO    &       & Ret   & SR    & TO    &       & Ret   & SR    & TO    &       & Ret   & SR    & TO    &       & Ret   & SR    & TO \\
%\cmidrule{1-4}\cmidrule{6-8}\cmidrule{10-12}\cmidrule{14-16}\cmidrule{18-20}
\cmidrule{1-20}
Original & 0.00  & 0.01  & 281\% &       & 0.05  & 0.18  & 805\% &       & 0.06  & 0.26  & 1604\% &       & 0.31  & 1.54  & 2480\% &       & 0.39  & 2.60  & 3018\% \\
\cmidrule{2-16}\cmidrule{18-20}          & \multicolumn{19}{c}{Panel A:  TWMA-RN} \\
\cmidrule{2-20}    RN18-D5/R1 & 0.00  & 0.00  & 148\% &       & 0.16  & 0.35  & 688\% &       & 0.09  & 0.35  & 1457\% &       & 0.26  & 1.22  & 2149\% &       & 0.42  & 2.77  & 2288\% \\
    RN18-D20/R1 & 0.01  & 0.01  & 233\% &       & 0.14  & 0.42  & 692\% &       & 0.12  & 0.44  & 1289\% &       & 0.28  & 1.21  & 2254\% &       & 0.40  & 2.58  & 2733\% \\
    RN18-D60/R1 & 0.01  & 0.01  & 199\% &       & 0.12  & 0.39  & 505\% &       & 0.09  & 0.38  & 1593\% &       & 0.27  & 1.14  & 2537\% &       & 0.38  & 2.60  & 2547\% \\
\cmidrule{2-20}    RN34-D5/R1 & \pmb{0.04} & \pmb{0.17} & \pmb{183\%} &       & \pmb{0.22} & \pmb{0.68} & \pmb{509\%} &       & \pmb{0.18} & \pmb{0.83} & \pmb{1177\%} &       & \pmb{0.33} & \pmb{1.96} & \pmb{1674\%} &       & \pmb{0.58} & \pmb{4.51} & \pmb{2389\%} \\
    RN34-D20/R1 & 0.01  & 0.04  & 169\% &       & 0.16  & 0.55  & 533\% &       & 0.17  & 0.74  & 1109\% &       & 0.20  & 1.75  & 1993\% &       & 0.52  & 3.44  & 2478\% \\
    RN34-D60/R1 & 0.00  & 0.02  & 142\% &       & 0.14  & 0.52  & 494\% &       & 0.15  & 0.58  & 658\% &       & 0.15  & 1.47  & 1432\% &       & 0.48  & 3.21  & 2153\% \\
\cmidrule{2-20}    RN50-D5/R1 & 0.02  & 0.16  & 196\% &       & 0.20  & 0.67  & 503\% &       & 0.14  & 0.63  & 1589\% &       & 0.30  & 1.78  & 1738\% &       & 0.49  & 4.11  & 2647\% \\
    RN50-D20/R1 & 0.01  & 0.03  & 178\% &       & 0.18  & 0.54  & 527\% &       & 0.05  & 0.28  & 1689\% &       & 0.19  & 1.62  & 1433\% &       & 0.47  & 3.16  & 2241\% \\
    RN50-D60/R1 & 0.00  & 0.00  & 192\% &       & 0.13  & 0.49  & 497\% &       & 0.05  & 0.21  & 1577\% &       & 0.17  & 1.28  & 1524\% &       & 0.41  & 2.56  & 2148\% \\
\cmidrule{2-20}          & \multicolumn{19}{c}{Panel B:  TWMA-CN} \\
\cmidrule{2-20}    CN-D5/R1 & 0.02  & 0.04  & 206\% &       & 0.06  & 0.23  & 649\% &       & 0.09  & 0.30  & 1221\% &       & 0.17  & 1.32  & 1834\% &       & 0.27  & 1.91  & 2833\% \\
    CN-D20/R1 & 0.01  & 0.02  & 184\% &       & 0.04  & 0.22  & 703\% &       & 0.06  & 0.22  & 1032\% &       & 0.14  & 1.17  & 1792\% &       & 0.20  & 1.53  & 2749\% \\
    CN-D60/R1 & 0.00  & -0.01  & 233\% &       & -0.01  & -0.07  & 688\% &       & 0.04  & 0.15  & 908\% &       & 0.09  & 0.95  & 1978\% &       & 0.23  & 1.71  & 2452\% \\
\cmidrule{2-20}          & \multicolumn{19}{c}{Panel C:  EWMA} \\
\cmidrule{2-20}    EWMA98  & 0.01  & 0.04  & 169\% &       & 0.06  & 0.21  & 487\% &       & 0.11  & 0.51  & 1080\% &       & 0.24  & 1.07  & 1679\% &       & 0.32  & 2.55  & 2148\% \\
    EWMA94  & 0.01  & 0.04  & 171\% &       & 0.06  & 0.21  & 494\% &       & 0.11  & 0.51  & 1091\% &       & 0.25  & 1.11  & 1702\% &       & 0.31  & 2.50  & 2194\% \\
    EWMA90   & 0.01  & 0.04  & 173\% &       & 0.06  & 0.21  & 499\% &       & 0.11  & 0.51  & 1101\% &       & 0.26  & 1.14  & 1722\% &       & 0.31  & 2.49  & 2278\% \\
%    EWMA(0.7)   & 0.01  & 0.03  & 186\% &       & 0.05  & 0.19  & 535\% &       & 0.11  & 0.53  & 1160\% &       & 0.28  & 1.27  & 1833\% &       & 0.24  & 2.35  & 2439\% \\
%    EWMA(0.5)   & 0.01  & 0.02  & 203\% &       & 0.05  & 0.18  & 582\% &       & 0.11  & 0.51  & 1237\% &       & 0.24  & 1.03  & 1967\% &       & 0.25  & 2.27  & 2477\% \\
    \bottomrule
    \end{tabular}%
    }
  \label{tab:daily_TWMA}%
  \begin{tablenotes}
  \item {Note: This table presents the performance metrics of out-of-sample one-day long-short portfolios proposed by different methods. For each class of trading signal, the portfolios are proposed by the methods based on original trading signal, TWMA-RN-based trading signal (in Panel A), TWMA-CN-based trading signal (in Panel B), and EWMA-based trading signal (in Panel C); and the performance of all portfolios is measured by three metrics: annualized excess return (Ret), annualized Sharpe ratio (SR), and monthly turnover (TO). Here, in each class of trading signal, the values of Ret, SR, and TO are in boldface if the related portfolio has the highest value of SR.}
  \end{tablenotes}
\end{table}%

\begin{table}[!h]
  \centering
  \caption{One-week long-short portfolio performance across various methods.}
  \scalebox{0.82}{
    \begin{tabular}{ccccrcccccccrccccccc}
    \toprule
          & \multicolumn{3}{c}{MOM} &       & \multicolumn{3}{c}{STR} &       & \multicolumn{3}{c}{WSTR} &       & \multicolumn{3}{c}{TREND} &       & \multicolumn{3}{c}{ALPHA} \\
\cmidrule{2-4}\cmidrule{6-8}\cmidrule{10-12}\cmidrule{14-16}\cmidrule{18-20}    Method & Ret   & SR    & TO    &       & Ret   & SR    & TO    &       & Ret   & SR    & TO    &       & Ret   & SR    & TO    &       & Ret   & SR    & TO \\
%\cmidrule{1-4}\cmidrule{6-8}\cmidrule{10-12}\cmidrule{14-16}\cmidrule{18-20}
\cmidrule{1-20}
Original & 0.02  & 0.07  & 129\% &       & 0.12  & 0.46  & 357\% &       & 0.12  & 0.63  & 682\% &       & 0.27  & 1.11  & 412\% &       & 0.28  & 1.77  & 582\% \\
\cmidrule{2-16}\cmidrule{18-20}          & \multicolumn{19}{c}{Panel A:  TWMA-RN} \\
\cmidrule{2-20}    RN18-D5/R5 & 0.02  & 0.03  & 117\% &       & 0.10  & 0.27  & 218\% &       & 0.09  & 0.41  & 428\% &       & 0.11  & 0.78  & 198\% &       & 0.27  & 1.61  & 445\% \\
    RN18-D20/R5 & 0.03  & 0.06  & 125\% &       & 0.14  & 0.45  & 325\% &       & 0.13  & 0.62  & 581\% &       & 0.14  & 0.86  & 211\% &       & 0.31  & 1.68  & 401\% \\
    RN18-D60/R5 & 0.03  & 0.05  & 119\% &       & 0.12  & 0.36  & 295\% &       & 0.04  & 0.29  & 396\% &       & 0.12  & 0.82  & 206\% &       & 0.22  & 0.87  & 430\% \\
\cmidrule{2-20}    RN34-D5/R5 & 0.03  & 0.07  & 113\% &       & 0.16  & 0.59  & 319\% &       & 0.15  & 0.64  & 628\% &       & 0.17  & 1.16  & 325\% &       & 0.35  & 1.64  & 474\% \\
    RN34-D20/R5 & \pmb{0.04} & \pmb{0.12} & \pmb{175\%} &       & \pmb{0.21} & \pmb{0.76} & \pmb{241\%} &       & \pmb{0.23} & \pmb{0.79} & \pmb{408\%} &       & \pmb{0.18} & \pmb{1.23} & \pmb{203\%} &       & 0.35  & 1.76  & 412\% \\
    RN34-D60/R5 & 0.01  & 0.03  & 85\%  &       & 0.10  & 0.36  & 330\% &       & 0.07  & 0.33  & 405\% &       & 0.08  & 0.49  & 239\% &       & 0.22  & 0.89  & 426\% \\
\cmidrule{2-20}    RN50-D5/R5 & 0.02  & 0.06  & 104\% &       & 0.14  & 0.54  & 326\% &       & 0.11  & 0.53  & 567\% &       & 0.16  & 1.14  & 361\% &       & 0.31  & 1.59  & 448\% \\
    RN50-D20/R5 & 0.03  & 0.09  & 137\% &       & 0.18  & 0.66  & 218\% &       & 0.22  & 0.69  & 534\% &       & 0.17  & 1.13  & 284\% &       & 0.33  & 1.63  & 397\% \\
    RN50-D60/R5 & 0.02  & 0.03  & 79\%  &       & 0.11  & 0.38  & 347\% &       & 0.05  & 0.29  & 692\% &       & 0.07  & 1.09  & 295\% &       & 0.19  & 0.82  & 407\% \\
\cmidrule{2-20}          & \multicolumn{19}{c}{Panel B:  TWMA-CN} \\
\cmidrule{2-20}    CN-D5/R5 & 0.00  & -0.02  & 112\% &       & 0.10  & 0.35  & 291\% &       & 0.10  & 0.45  & 438\% &       & 0.07  & 0.69  & 418\% &       & 0.23  & 0.97  & 522\% \\
    CN-D20/R5 & 0.01  & 0.03  & 89\%  &       & 0.11  & 0.38  & 284\% &       & 0.13  & 0.50  & 369\% &       & 0.13  & 0.80  & 321\% &       & 0.20  & 0.95  & 599\% \\
    CN-D60/R5 & 0.00  & -0.27  & 94\%  &       & -0.02  & -0.18  & 215\% &       & 0.04  & 0.30  & 372\% &       & 0.09  & 0.63  & 362\% &       & 0.13  & 0.57  & 563\% \\
\cmidrule{2-20}          & \multicolumn{19}{c}{Panel C:  EWMA} \\
\cmidrule{2-20}    EWMA98 & 0.01  & 0.02  & 112\% &       & 0.13  & 0.46  & 319\% &       & 0.11  & 0.52  & 633\% &       & 0.18  & 0.68  & 329\% &       & 0.26  & 1.84  & 551\% \\
    EWMA94 & 0.01  & 0.02  & 113\% &       & 0.12  & 0.46  & 320\% &       & 0.11  & 0.54  & 635\% &       & 0.18  & 0.69  & 332\% &       & 0.26  & 1.84  & 547\% \\
    EWMA90 & 0.01  & 0.03  & 113\% &       & 0.13  & 0.47  & 322\% &       & 0.12  & 0.57  & 637\% &       & 0.18  & 0.69  & 334\% &       & \pmb{0.29} & \pmb{2.03} & \pmb{542\%} \\
%    EWMA(0.7) & 0.00  & 0.01  & 116\% &       & 0.13  & 0.48  & 328\% &       & 0.13  & 0.62  & 647\% &       & 0.20  & 0.76  & 346\% &       & 0.28  & 2.02  & 538\% \\
%    EWMA(0.5) & 0.01  & 0.02  & 119\% &       & 0.12  & 0.45  & 335\% &       & 0.14  & 0.66  & 657\% &       & 0.22  & 0.87  & 361\% &       & 0.28  & 1.99  & 522\% \\
    \bottomrule
    \end{tabular}%
    }
    \begin{tablenotes}
  \item {Note: This table presents the performance metrics of out-of-sample one-week long-short portfolios proposed by different methods. Other descriptions are consistent with those in Table \ref{tab:daily_TWMA}.}
  \end{tablenotes}
  \label{tab:weekly_TWMA}%
\end{table}%

\subsubsection{Impact of Image Structure}

\begin{figure}
    \centering
    \includegraphics[width = 0.9\linewidth]{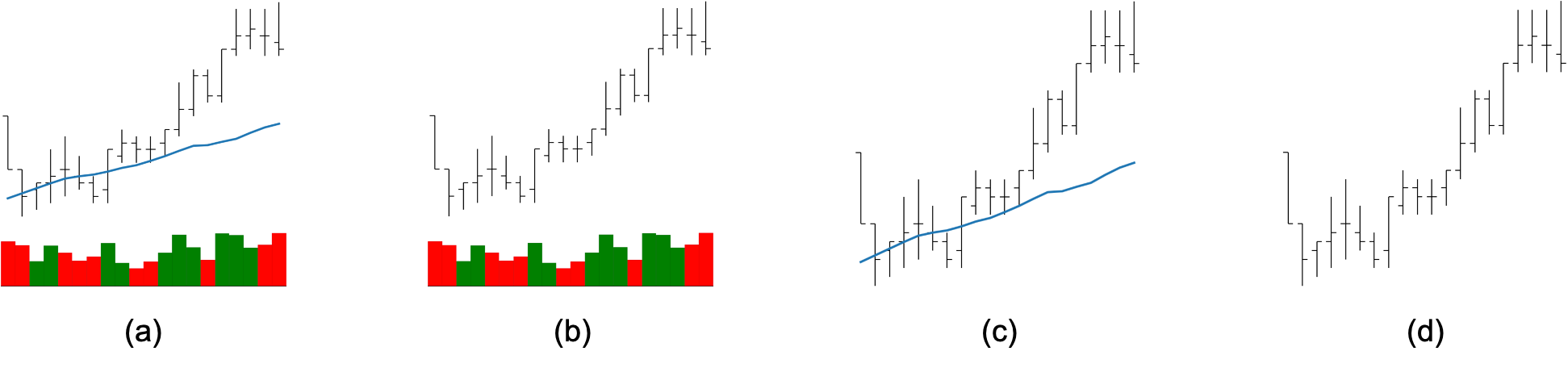}
    \caption{(Color online) Different types of 20-day image, constructed by (a) OHLC bars + volume bars + moving average price line, (b) OHLC bars + volume bars, (c) OHLC bars + moving average price line, and (d) OHLC bars only.}
    \label{fig:image_choice}
\end{figure}

To implement the TWMA method, our $D$-day images chart historical stock information through OHLC bars, volume bars, and moving average price lines. Then, an interesting question arises: Which charting part in the image is more informative for the TWMA method? To answer this question, we examine the performance of portfolios generated from three RN34 methods under four different types of image, which are constructed by (a) OHLC bars + volume bars + moving average price line, (b) OHLC bars + volume bars, (c) OHLC bars + moving average price line, and (d) OHLC bars only. Figure \ref{fig:image_choice} gives a visual illustration of these four types of image, where the first type image is the benchmark one analyzed above.

Under four different types of image, Table \ref{tab:daily_image_choice} reports the values of annualized excess return, annualized Sharpe ratio, and monthly turnover for all proposed portfolios, which target the one-day-ahead return predictions. By comparing the results of Panels A and B in Table \ref{tab:daily_image_choice}, we find that excluding the moving average price lines increases the value of Sharpe ratio only for the D5 case in the classes of STR and WSTR. This finding may be attributed to the fact that the 5-day moving average price line provides little information for identifying long-term reference points and negatively impacts the recognition of reversal patterns in $5$-day images.

Next, the comparison between Panels A and C in Table \ref{tab:daily_image_choice} indicates that the volume bars are crucial for the TWMA method to identify key important regions. This is because when including volume bars, the values of Sharpe ratio always become higher, except for the RN34-D20/R1 method in the class of MOM. Particularly, the inclusion of volume bars makes the largest increase on the value of Sharpe ratio (from 0.03 to 0.17) for the RN34-D5/R1 method in the class of MOM, and it generally leads to similar increasing rate of Sharpe ratio across the choice of $D$ in the classes of STR, WSTR, TREND, and ALPHA. These interesting findings illustrate the importance of incorporating multiple sources of information, since the MOM, STR, WSTR, and TREND signals are constructed without considering volume-related information, yet trading volume data enhances these signals for portfolio selections. This observation could also present a fact that the important regions on volume bars are temporarily correlated with those on OHLC bars and moving average price lines, making the TWMA method more efficient.

Moreover, as evidenced by the comparison between Panels A and D in Table \ref{tab:daily_image_choice}, the OHLC bars are much more important than the volume bars and moving average lines in the classes of STR, WSTR, TREND, and ALPHA. This conclusion is drawn from the observation that in most cases of these four classes, the proportion of two Sharpe ratio values in Panels A and D exceeds 70\%. For the class of MOM, none of the charting parts seems informative, except for the case of D5 in which the volume bars are more important than the OHLC bars. Finally, we can see that the values of differenced Sharpe ratio between Panels C and D are less significant than those between Panels B and D. This suggests that the moving average lines are generally less informative than the volume bars.

Likewise, Table \ref{tab:weekly_image_choice} further assesses the performance of one-week portfolios, which target the five-day-ahead returns. Broadly speaking, the majority of findings from this table are consistent with those from Table \ref{tab:daily_image_choice}; they emphasize the importance of using the three charting parts altogether, with the OHLC bars having the highest priority and the moving average price lines having the lowest priority.

\begin{table}[!h]
  \centering
  \caption{One-day long-short portfolio performance across different image structures.}
  \scalebox{0.8}{
    \begin{tabular}{ccccrcccccccrccccccc}
    \toprule
          & \multicolumn{3}{c}{MOM} &       & \multicolumn{3}{c}{STR} &       & \multicolumn{3}{c}{WSTR} &       & \multicolumn{3}{c}{TREND} &       & \multicolumn{3}{c}{ALPHA} \\
\cmidrule{2-4}\cmidrule{6-8}\cmidrule{10-12}\cmidrule{14-16}\cmidrule{18-20}    Method & Ret   & SR    & TO    &       & Ret   & SR    & TO    &       & Ret   & SR    & TO    &       & Ret   & SR    & TO    &       & Ret   & SR    & TO \\
%\cmidrule{1-4}\cmidrule{6-8}\cmidrule{10-12}\cmidrule{14-16}\cmidrule{18-20}
\cmidrule{1-20}
%Original & 0.00  & 0.01  & 281\% &       & 0.05  & 0.18  & 805\% &       & 0.06  & 0.26  & 1604\% &       & 0.31  & 1.54  & 2480\% &       & 0.39  & 2.60  & 3018\% \\
\cmidrule{2-16}\cmidrule{18-20}          & \multicolumn{19}{c}{Panel A:  OHLC + VB + MA} \\
\cmidrule{2-20}    RN34-D5/R1 & \pmb{0.04} & \pmb{0.17} & \pmb{183\%} &       & 0.22  & 0.68  & 509\% &       & 0.18  & 0.83  & 1177\% &       & \pmb{0.33} & \pmb{1.96} & \pmb{1674\%} &       & \pmb{0.58} & \pmb{4.51} & \pmb{2389\%} \\
    RN34-D20/R1 & 0.01  & 0.04  & 169\% &       & 0.16  & 0.55  & 533\% &       & 0.17  & 0.74  & 1109\% &       & 0.20  & 1.75  & 1993\% &       & 0.52  & 3.44  & 2478\% \\
    RN34-D60/R1 & 0.00  & 0.02  & 142\% &       & 0.14  & 0.52  & 494\% &       & 0.15  & 0.58  & 658\% &       & 0.15  & 1.47  & 1432\% &       & 0.48  & 3.21  & 2153\% \\
\cmidrule{2-20}          & \multicolumn{19}{c}{Panel B:  OHLC + VB} \\
\cmidrule{2-20}    RN34-D5/R1 & 0.03  & 0.12  & 209\% &       & \pmb{0.21} & \pmb{0.77} & \pmb{993\%} &       & \pmb{0.25} & \pmb{0.97} & \pmb{1787\%} &       & 0.21  & 1.68  & 1396\% &       & 0.29  & 3.54  & 3151\% \\
    RN34-D20/R1 & 0.01  & 0.04  & 115\% &       & 0.09  & 0.32  & 672\% &       & 0.16  & 0.73  & 1019\% &       & 0.18  & 1.43  & 1132\% &       & 0.24  & 2.85  & 3067\% \\
    RN34-D60/R1 & 0.00  & 0.00  & 161\% &       & 0.06  & 0.22  & 4.56  &       & 0.11  & 0.53  & 894\% &       & 0.13  & 1.11  & 1413\% &       & 0.20  & 2.27  & 3129\% \\
\cmidrule{2-20}          & \multicolumn{19}{c}{Panel C:  OHLC + MA} \\
\cmidrule{2-20}    RN34-D5/R1 & 0.01  & 0.03  & 159\% &       & 0.16  & 0.64  & 454\% &       & 0.15  & 0.59  & 1333\% &       & 0.21  & 1.44  & 1461\% &       & 0.24  & 3.14  & 3307\% \\
    RN34-D20/R1 & 0.01  & 0.04  & 195\% &       & 0.14  & 0.50  & 623\% &       & 0.12  & 0.49  & 1360\% &       & 0.22  & 1.40  & 1140\% &       & 0.25  & 2.78  & 3463\% \\
    RN34-D60/R1 & -0.01  & -0.04  & 263\% &       & 0.10  & 0.40  & 1035\% &       & 0.04  & 0.14  & 1278\% &       & 0.12  & 1.28  & 955\% &       & 0.15  & 1.89  & 3420\% \\
\cmidrule{2-20}          & \multicolumn{19}{c}{Panel D:  OHLC} \\
\cmidrule{2-20}    RN34-D5/R1 & 0.01  & 0.02  & 162\% &       & 0.16  & 0.64  & 439\% &       & 0.15  & 0.60  & 1301\% &       & 0.21  & 1.40  & 1299\% &       & 0.25  & 3.15  & 3299\% \\
    RN34-D20/R1 & 0.01  & 0.02  & 192\% &       & 0.13  & 0.48  & 661\% &       & 0.12  & 0.49  & 1352\% &       & 0.20  & 1.40  & 1153\% &       & 0.23  & 2.69  & 3311\% \\
    RN34-D60/R1 & 0.00  & 0.00  & 204\% &       & 0.08  & 0.39  & 597\% &       & 0.04  & 0.14  & 1280\% &       & 0.11  & 1.12  & 957\% &       & 0.14  & 1.84  & 3652\% \\
    \bottomrule
    \end{tabular}%
    }
  \label{tab:daily_image_choice}%
   \begin{tablenotes}
    \item {Note: This table presents the performance metrics of out-of-sample one-day long-short portfolios proposed by three different RN34 methods, where each RN34 method is implemented using the images with four different structures as indicated in  Panels A, B, C, and D. Here, ``OHLC'' indicates that the OHLC bars are included in the image, whereas ``VB'' and ``MA'' denote the inclusion of volume bars and the moving average line in the image, respectively. Other descriptions inherit from Table \ref{tab:daily_TWMA}.
    }
    \end{tablenotes}
\end{table}%

\begin{table}[!h]
  \centering
  \caption{One-week long-short portfolio performance across different image structures.}
  \scalebox{0.82}{
    \begin{tabular}{cccccccccccccccccccc}
    \toprule
          & \multicolumn{3}{c}{MOM} &       & \multicolumn{3}{c}{STR} &       & \multicolumn{3}{c}{WSTR} &       & \multicolumn{3}{c}{TREND} &       & \multicolumn{3}{c}{ALPHA} \\
\cmidrule{2-4}\cmidrule{6-8}\cmidrule{10-12}\cmidrule{14-16}\cmidrule{18-20}    Method & Ret   & SR    & TO    &       & Ret   & SR    & TO    &       & Ret   & SR    & TO    &       & Ret   & SR    & TO    &       & Ret   & SR    & TO \\
%\cmidrule{1-4}\cmidrule{6-8}\cmidrule{10-12}\cmidrule{14-16}\cmidrule{18-20}
\cmidrule{1-20}
%Original & 0.02  & 0.07  & 129\% &       & 0.12  & 0.46  & 357\% &       & 0.12  & 0.63  & 682\% &       & 0.27  & 1.11  & 412\% &       & \textbf{0.28 } & \textbf{1.77 } & \textbf{582\%} \\
\cmidrule{2-16}\cmidrule{18-20}          & \multicolumn{19}{c}{Panel A:  OHLC + VB + MA} \\
\cmidrule{2-20}    RN34-D5/R5 & 0.03  & 0.07  & 113\% &       & 0.16  & 0.59  & 319\% &       & 0.15  & 0.64  & 628\% &       & 0.17  & 1.16  & 325\% &       & 0.35  & 1.64  & 474\% \\
   RN34- D20/R5 & \pmb{0.04} & \pmb{0.12} & \pmb{175\%} &       & \pmb{0.21} & \pmb{0.76} & \pmb{241\%} &       & \pmb{0.23} & \pmb{0.79} & \pmb{408\%} &       & \pmb{0.18} & \pmb{1.23} & \pmb{203\%} &       & \pmb{0.35}  & \pmb{1.76}  & \pmb{412\%} \\
    RN34-D60/R5 & 0.01  & 0.03  & 85\%  &       & 0.10  & 0.36  & 330\% &       & 0.07  & 0.33  & 405\% &       & 0.08  & 0.49  & 239\% &       & 0.22  & 0.89  & 426\% \\
\cmidrule{2-20}          & \multicolumn{19}{c}{Panel B:  OHLC + VB} \\
\cmidrule{2-20}    RN34-D5/R5 & 0.00  & 0.00  & 115\% &       & 0.12  & 0.44  & 325\% &       & 0.09  & 0.41  & 406\% &       & 0.15  & 1.07  & 334\% &       & 0.28  & 1.46  & 709\% \\
    RN34-D20/R5 & 0.02  & 0.05  & 76\%  &       & 0.19  & 0.66  & 338\% &       & 0.13  & 0.64  & 641\% &       & 0.17  & 0.91  & 204\% &       & 0.26  & 1.29  & 623\% \\
    RN34-D60/R5 & -0.02  & -0.07  & 101\% &       & 0.15  & 0.52  & 238\% &       & 0.17  & 0.53  & 420\% &       & 0.09  & 0.70  & 255\% &       & 0.21  & 0.82  & 586\% \\
\cmidrule{2-20}          & \multicolumn{19}{c}{Panel C:  OHLC + MA} \\
\cmidrule{2-20}    RN34-D5/R5 & 0.01  & 0.02  & 112\% &       & 0.08  & 0.29  & 317\% &       & 0.07  & 0.36  & 624\% &       & 0.12  & 1.05  & 319\% &       & 0.24  & 1.29  & 678\% \\
    RN34-D20/R5 & 0.03  & 0.07  & 73\%  &       & 0.14  & 0.49  & 232\% &       & 0.11  & 0.66  & 386\% &       & 0.09  & 0.93  & 493\% &       & 0.23  & 1.15  & 669\% \\
    RN34-D60/R5 & -0.01  & -0.04  & 77\%  &       & 0.10  & 0.48  & 286\% &       & 0.08  & 0.48  & 394\% &       & 0.08  & 0.87  & 214\% &       & 0.18  & 0.78  & 603\% \\
\cmidrule{2-20}          & \multicolumn{19}{c}{Panel D:  OHLC} \\
\cmidrule{2-20}    RN34-D5/R5 & 0.01  & 0.03  & 111\% &       & 0.08  & 0.30  & 316\% &       & 0.05  & 0.26  & 617\% &       & 0.11  & 0.92  & 328\% &       & 0.18  & 1.25  & 582\% \\
    RN34-D20/R5 & 0.01  & 0.03  & 73\%  &       & 0.12  & 0.40  & 234\% &       & 0.06  & 0.24  & 403\% &       & 0.08  & 0.77  & 390\% &       & 0.27  & 1.31  & 472\% \\
    RN34-D60/R5 & 0.02  & 0.08  & 59\%  &       & 0.20  & 0.71  & 212\% &       & 0.20  & 0.70  & 351\% &       & 0.10  & 0.74  & 360\% &       & 0.22  & 1.00  & 493\% \\
    \bottomrule
    \end{tabular}%
    }
    \begin{tablenotes}
    \item {Note: This table presents the performance metrics of out-of-sample one-week long-short portfolios proposed by three different RN34 methods. Other descriptions are the same as those in Table \ref{tab:daily_image_choice}.}
    \end{tablenotes}
  \label{tab:weekly_image_choice}%
\end{table}%

\subsubsection{Impact of Stock Size}\label{sec:stock_size}

To avoid abnormal price patterns associated with small-cap stocks, we restrict our previous analysis to the largest $1,800$ stocks in the Chinese stock market. However, this size filter reduces the number of samples, and may omit valuable patterns that are inherent in small-cap stocks. To investigate the effect of stock size on the TWMA method, we collect daily stock data for all firms listed on the Shanghai and Shenzhen Stock Exchanges, and consider three groups of stocks for model training and portfolio backtesting: (a) large-cap stocks (largest 1,800 stocks), (b) small-cap stocks (stocks not included in the large-cap group), and (c) all stocks. We re-categorize these groups every six months by market capitalization and generate the images and labels to build the dataset of each group. Notably, there are no overlapping samples between the large-cap and small-cap datasets, while the dataset for all stocks is the union of the large-cap and small-cap datasets.

Based on the grouping above, we then consider 9 different configurations from the combinations of training and backtesting: train the ResNet model on each group dataset of training and validation samples separately, and use the well-trained model to do out-of-sample portfolio backtesting on each group dataset of testing samples. Since our aim is to probe how the stock size affects the performance of the TWMA method, we only focus on the ResNet34 model, which performs the best in portfolio selections as demonstrated above. Specifically, we train the ResNet34 model based on 5-day images for targeting one-day holding period returns, and then take the trained model with the RN34-D5/R1 method to propose out-of-sample portfolios for backtesting. For targeting one-week holding period returns, we apply the same procedure, except for the replacement of RN34-D5/R1 method with RN34-D20/R5 method. For brevity, we refer the RN34 method from the trained ResNet34 model on group of large-cap (or small-cap or all) stocks to as the large-cap (or small-cap or full) RN34 method. As a benchmark, we also account for 3 more different configurations, each of which uses the same group of stocks for training and backtesting based on the original method. Here, the training implementation is needed only for proposing portfolios in the classes of TREND and ALPHA.

Under different configurations, Table \ref{tab:daily_size} displays the performance of all proposed portfolios, which target one-day-ahead returns.
From the Panel A of Table \ref{tab:daily_size}, we find that regardless of the signal class, the large-cap RN34 method performs the best with the value of Sharpe ratio ranging from 0.17 to 4.51. In this case, the full RN34 method is the second best, exhibiting a minor improvement over the small-cap RN34 method. These findings suggest that when predicting the price trend of large-cap stocks, the identification of general price patterns from the images of large-cap stocks proves to be more precise than those from small-cap stocks; however, incorporating the images of all stocks could potentially impair the data-mining efficacy of the TWMA method.

Next, we observe from the Panel B of Table \ref{tab:daily_size} that when predicting the price trend of small-cap stocks, there is no dominating performance of the large-cap RN34 method over others. For instance, the small-cap and full RN34 methods perform the best in the classes of STR and TREND, respectively, although their advantage over the large-cap RN method is not evident. The above observation is not unexpected, since the small-cap stocks could have some abnormal price patterns. It appears that these abnormal price patterns have varying degrees of correlation with different trading signals, and they are hard to detect if the training procedure does not include the images of small-cap stocks.

Moreover, the results of Panel C of Table \ref{tab:daily_size} reveal that the advantage of large-cap RN34 method over others remains for
predicting the price trend of all stocks. In this case, the small-cap and full RN34 methods have the comparable performance, with no clear dominance of one over the other. Finally, we highlight that regardless of the way of configuration, all of RN34 methods perform much better than the original method, with higher values of Sharpe ratio and lower values of turnover. This demonstrates the notable effectiveness of the TWMA method from the viewpoint of robustness.

Notably, most of above findings appear in Table \ref{tab:weekly_size}, which gives the results for one-week portfolios.
A noteworthy exception is that when predicting the price trend of small-cap stocks, the small-cap RN34 method performs slightly better than the large-cap RN34 method. This is probably because the small-cap stocks have more unique abnormal price patterns, which are valuable for their own predictions with longer horizons. In summary, we conclude that the large-cap RN34 method has a desirable and robust performance, while the small-cap or full RN34 method could show certain advantage in the long-horizon prediction of small-cap stocks.

\begin{table}[!h]
  \centering
  \caption{One-day long-short portfolio performance across different training and backtesting configurations.}
  \scalebox{0.71}{
    \begin{tabular}{cccccrcccccccrccccccc}
    \toprule
          & & \multicolumn{3}{c}{MOM} &       & \multicolumn{3}{c}{STR} &       & \multicolumn{3}{c}{WSTR} &       & \multicolumn{3}{c}{TREND} &       & \multicolumn{3}{c}{ALPHA} \\
\cmidrule{3-5}\cmidrule{7-9}\cmidrule{11-13}\cmidrule{15-17}\cmidrule{19-21}
Method & Group & Ret   & SR    & TO    &       & Ret   & SR    & TO    &       & Ret   & SR    & TO    &       & Ret   & SR    & TO    &       & Ret   & SR    & TO \\
\midrule
          & & \multicolumn{19}{c}{Panel A: Backtesting on group of large-cap stocks} \\
\cmidrule{3-21}
Original & Large-cap stocks & 0.00  & 0.01  & 281\% &       & 0.05  & 0.18  & 805\% &       & 0.06  & 0.26  & 1604\% &       & 0.31  & 1.54  & 2480\% &       & 0.39  & 2.60  & 3018\% \\
    RN34-D5/R1  & Large-cap stocks & \pmb{0.04} & \pmb{0.17} & \pmb{183\%} &       & \pmb{0.22 } & \pmb{0.68} & \pmb{509\%} &       & \pmb{0.18} & \pmb{0.83} & \pmb{1177\%} &       & \pmb{0.33} & \pmb{1.96} & \pmb{1674\%} &       & \pmb{0.58} & \pmb{4.51} & \pmb{2389\%} \\
     & Small-cap stocks & 0.01  & 0.06  & 205\% &       & 0.14  & 0.49  & 1096\% &       & 0.11  & 0.53  & 1034\% &       & 0.24  & 1.52  & 1502\% &       & 0.25  & 3.28  & 2783\% \\
       & All stocks & 0.02  & 0.08  & 190\% &       & 0.21  & 0.62  & 1131\% &       & 0.16  & 0.73  & 1412\% &       & 0.26  & 1.65  & 1496\% &       & 0.25  & 3.30  & 2406\% \\
\cmidrule{3-21}
& & \multicolumn{19}{c}{Panel B: Backtesting on group of small-cap stocks} \\
\cmidrule{3-21}
Original & Small-cap stocks & -0.10  & -0.72  & 318\% &       & 0.15  & 0.78  & 815\% &       & 0.20  & 1.00  & 1554\% &       & 0.37  & 2.72  & 2163\% &       & 0.32  & 3.94  & 3295\% \\
    RN34-D5/R1 & Large-cap stocks & \pmb{-0.04} & \pmb{-0.18} & \pmb{190\%} &       & 0.27  & 1.29  & 515\% &       & \pmb{0.31} & \pmb{1.70} & \pmb{1153\%} &       & 0.37  & 2.90  & 1609\% &       & \pmb{0.33} & \pmb{4.21} & \pmb{3409\%} \\
     & Small-cap stocks & -0.13  & -0.77  & 175\% &       & \pmb{0.26} & \pmb{1.40} & \pmb{1080\%} &       & 0.28  & 1.33  & 1694\% &       & 0.50  & 2.84  & 1447\% &       & 0.34  & 4.10  & 3225\% \\
       & All stocks & -0.12  & -0.72  & 174\% &       & 0.21  & 1.00  & 476\% &       & 0.24  & 1.29  & 1286\% &       & \pmb{0.41} & \pmb{2.97} & \pmb{1434\%} &       & 0.32  & 4.14  & 3310\% \\
\cmidrule{3-21}
& & \multicolumn{19}{c}{Panel C: Backtesting on group of all stocks} \\
\cmidrule{3-21}
Original & All stocks & \pmb{-0.07} & \pmb{-0.33} & \pmb{280\%} &       & 0.11  & 0.49  & 799\% &       & 0.05  & 0.25  & 1565\% &       & 0.36  & 2.36  & 2389\% &       & 0.34  & 3.16  & 3344\% \\
    RN34-D5/R1 & Large-cap stocks & -0.07  & -0.34  & 161\% &       & \pmb{0.27} & \pmb{1.06} & \pmb{479\%} &       & \pmb{0.24} & \pmb{1.31} & \pmb{1152\%} &       & \pmb{0.31} & \pmb{2.67} & \pmb{1621\%} &       & \pmb{0.32} & \pmb{4.01} & \pmb{3142\%} \\
     & Small-cap stocks & -0.08  & -0.40  & 163\% &       & 0.22  & 0.91  & 1071\% &       & 0.19  & 1.02  & 1682\% &       & 0.32  & 2.23  & 1455\% &       & 0.28  & 3.28  & 3286\% \\
       & All stocks & -0.07  & -0.37  & 162\% &       & 0.18  & 0.93  & 765\% &       & 0.12  & 0.67  & 1045\% &       & 0.33  & 2.58  & 1443\% &       & 0.26  & 3.80  & 3317\% \\
    \bottomrule
    \end{tabular}%
    }
    \begin{tablenotes}
    \item {Note: This table demonstrates the performance of out-of-sample one-day long-short backtesting portfolios, which are proposed on group of large-cap stocks (Panel A) or group of small-cap stocks (Panel B) or group of all stocks (Panel C) under either the RN34-D5/R1 method or original method.
    Here, the RN34-D5/R1 method uses the ResNet34 model trained on group of large-cap stocks or group of small-cap stocks or  group of all stocks, and the original method need the training implementation (on the same group of stocks as for the backtesting) only in the classes of TREND and ALPHA. For each class of trading signal in every panel, the values of Ret, SR, and TO are in boldface if the related portfolio has the highest value of SR. Other descriptions are consistent with those in Table \ref{tab:daily_TWMA}.}
    \end{tablenotes}
  \label{tab:daily_size}%
\end{table}%

   %sorted on the TWMA-based trading signals with distinct stock pools for model training and portfolio backtesting. Large-cap, small-cap and all stocks are the three groups mentioned in Section \ref{sec:stock_size}.

\begin{table}[!h]
  \centering
  \caption{One-week long-short portfolio performance across different training and backtesting configurations.}
  \scalebox{0.73}{
    \begin{tabular}{cccccrcccccccrccccccc}
    \toprule
          & & \multicolumn{3}{c}{MOM} &       & \multicolumn{3}{c}{STR} &       & \multicolumn{3}{c}{WSTR} &       & \multicolumn{3}{c}{TREND} &       & \multicolumn{3}{c}{ALPHA} \\
\cmidrule{3-5}\cmidrule{7-9}\cmidrule{11-13}\cmidrule{15-17}\cmidrule{19-21}
Method & Group & Ret   & SR    & TO    &       & Ret   & SR    & TO    &       & Ret   & SR    & TO    &       & Ret   & SR    & TO    &       & Ret   & SR    & TO \\
\midrule
          & & \multicolumn{19}{c}{Panel A: Backtesting on group of large-cap stocks} \\
\cmidrule{3-21}
Original & Large-cap stocks & 0.02  & 0.07  & 129\% &       & 0.12  & 0.46  & 357\% &       & 0.12  & 0.63  & 682\% &       & 0.27  & 1.11  & 412\% &       & \pmb{0.28} & \pmb{1.77} & \pmb{582\%} \\
     RN34-D20/R5 & Large-cap stocks & \pmb{0.04} & \pmb{0.12} & \pmb{175\%} &       & \pmb{0.21} & \pmb{0.76} & \pmb{241\%} &       & \pmb{0.23} & \pmb{0.79} & \pmb{408\%} &       & \pmb{0.18} & \pmb{1.23} & \pmb{203\%} &       & 0.35  & 1.76  & 412\% \\
     & Small-cap stocks & 0.02  & 0.06  & 74\%  &       & 0.14  & 0.50  & 234\% &       & 0.16  & 0.57  & 365\% &       & 0.22  & 1.20  & 213\% &       & 0.29  & 1.53  & 378\% \\
       & All stocks & 0.02  & 0.06  & 79\%  &       & 0.15  & 0.53  & 245\% &       & 0.13  & 0.58  & 418\% &       & 0.09  & 0.33  & 208\% &       & 0.24  & 1.72  & 465\% \\
\cmidrule{3-21}
   & & \multicolumn{19}{c}{Panel B:  Backtesting on group of small-cap stocks} \\
\cmidrule{3-21}
Original & Small-cap stocks & -0.12  & -0.90  & 142\% &       & 0.32  & 1.43  & 361\% &       & 0.36  & 1.50  & 673\% &       & 0.40  & 1.98  & 369\% &       & 0.20  & 2.08  & 681\% \\
     RN34-D20/R5 & Large-cap stocks & \pmb{-0.14} & \pmb{-0.78} & \pmb{80\%} &       & 0.32  & 1.51  & 242\% &       & \pmb{0.30} & \pmb{1.53} & \pmb{403\%} &       & 0.28  & 1.90  & 230\% &       & 0.22  & 2.37  & 703\% \\
     & Small-cap stocks & -0.15  & -0.86  & 79\%  &       & \pmb{0.34} & \pmb{1.64} & \pmb{235\%} &       & 0.27  & 1.36  & 364\% &       &
     \pmb{0.34} & \pmb{2.17} & \pmb{324\%} &       & \pmb{0.22} & \pmb{2.55} & \pmb{674\%} \\
       & All stocks & -0.14  & -0.78  & 83\%  &       & 0.32  & 1.59  & 245\% &       & 0.33  & 1.50  & 360\% &       & 0.29  & 2.03  & 317\% &       & 0.21  & 2.18  & 662\% \\
\cmidrule{3-21}
& & \multicolumn{19}{c}{Panel C:  Backtesting on group of all stocks} \\
\cmidrule{3-21}
Original & All stocks & -0.06  & -0.27  & 128\% &       & 0.18  & 0.81  & 357\% &       & 0.11  & 0.62  & 675\% &       & 0.34  & 1.61  & 401\% &       & 0.27  & 2.10  & 658\% \\
     RN34-D20/R5 & Large-cap stocks & \pmb{-0.07} & \pmb{-0.31} & \pmb{74\%} &       & \pmb{0.30} & \pmb{1.23} & \pmb{265\%} &       & \pmb{0.29} & \pmb{1.19} & \pmb{355\%} &       & \pmb{0.35} & \pmb{1.74} & \pmb{193\%} &       & 0.30  & 2.12  & 677\% \\
     & Small-cap stocks & -0.07  & -0.32  & 114\% &       & 0.28  & 1.16  & 231\% &       & 0.23  & 0.97  & 360\% &       & 0.21  & 1.17  & 200\% &       & 0.23  & 2.04  & 785\% \\
       & All stocks & -0.07  & -0.31  & 77\%  &       & 0.26  & 1.06  & 239\% &       & 0.06  & 0.28  & 412\% &       & 0.29  & 1.45  & 198\% &       & \pmb{0.26} & \pmb{2.14} & \pmb{704\%} \\
    \bottomrule
    \end{tabular}%
    }
    \begin{tablenotes}
    \item {Note: This table demonstrates the performance of out-of-sample one-week long-short backtesting portfolios, which are proposed on group of large-cap stocks (Panel A) or group of small-cap stocks (Panel B) or group of all stocks (Panel C)  under either the RN34-D20/R5 method or original method. Here, the RN34-D20/R5 method uses the ResNet34 model trained on group of large-cap stocks or group of small-cap stocks or  group of all stocks. Other descriptions inherit from Table \ref{tab:daily_size}. }
    \end{tablenotes}
  \label{tab:weekly_size}%
\end{table}%

\section{Time-scale Transfer Learning}\label{sec:time_transfer}

So far we have only applied the TWMA method to develop portfolios for predicting one- (or five-) day-ahead returns. To handle predictions over longer horizons, we need to design $D$-day images with a larger value of $D$. Yet, this could lead to two issues: First, the number of $D$-day image data samples becomes much smaller; Second, the price trend signal from $D$-day images is likely to be less impotent (\citealp{jiang2023re}).

To solve these issues, a possible solution is the time-scale transfer learning. For brevity, we show how to implement this transfer method for proposing one-month long-short portfolios, which target $20$-day-ahead returns. First, we down-sample the price and volume data from once per day to once every four days. After down-sampling, every 4-day data are collapsed into a one-period data, including the opening, high, low, and closing prices within this period, as well as the trading volume summed over this period. Then, we construct the five-period image (like the previous 5-day image), which draws the OHLC bars, volume bars, and moving average price lines for every five periods. Next, using  a pre-trained ResNet34 model in the  RN34-D5/R5 method, we input each five-period image on the testing sample to obtain its triple-I weights. Since each triple-I weight is for a period of four days, one-quarter of its value is assigned as the triple-I weight to each of the four days in this period. Consequently, we have the triple-I weights for 20 days within five periods of each input image, and we call them the transferred triple-I weights. Finally, our transfer method enhances the existing trading signals as done in (\ref{IWA_signal}), except that the triple-I weights are replaced with the transferred triple-I weights. Based on these enhanced trading signals, the long-short portfolios are proposed to predict 20-day ahead returns.

To assess the performance of the transfer method, we also consider its four competitors.
The first one is the original method, which constructs the portfolios from five original trading signals.
The second one is the baseline method, which proposes the RN34-D20/R20 portfolios from five TWMA-RN-based trading signals.
The third competitor is the re-train method. This re-train method is the same as the transfer method, except that it re-trains the ResNet34 model in the RN34-D5/R5 method using the collapsed five-period images to compute the triple-I weights. Clearly, the purpose of re-train method is to check the effectiveness of the down-sampling. The last competing method is the so-called ``baseline+transfer'' method, which computes the triple-I weights by averaging those from both baseline and transfer methods. This fusion method attempts to examine whether the triple-I weights could be more accurate through the combination of the information from two methods.

Table \ref{tab:transfer} presents an out-of-sample comparative analysis of one-month portfolios across five different methods.
From this table, we find that in terms of the maximized value of Sharpe ratio, the transfer method performs the best in the classes of MOM, STR, and WSTR, whereas it is slightly worse than the re-train method in the classes of TREND and ALPHA. This finding indicates that
the long-term momentum and short-term reversal patterns which unfold at high frequencies (e.g., 5-day) and low frequencies (e.g., 20-day)
are similar; however, certain complex patterns (used in the TREND and ALPHA signals) appear to depend on the data frequency, and
their information unfolding at low frequencies could be largely utilized by the down-sampling implementation.

Compared with the original method, it is evident that the baseline method has lower value of Sharpe ratio in each signal class, whereas
the transfer method has up to 40\% higher value of Sharpe ratio in five signal classes. Due to the poor performance of the baseline method, the ``baseline+tranfer'' method always performs worse than the transfer method. Hence, when targeting one-month-ahead or even longer horizon predictions, the transfer method (with either a pre-trained or re-trained model on images of high-frequency data) is more capable than the baseline method for detecting the general price patters from images of low-frequency data.

\begin{table}[ht]
  \centering
  \caption{One-month long-short portfolio performance across different learning methods.}
  \scalebox{0.82}{
    \begin{tabular}{ccccccccccccccccrccc}
    \toprule
          &  \multicolumn{3}{c}{MOM} &       & \multicolumn{3}{c}{STR} &       & \multicolumn{3}{c}{WSTR} &       & \multicolumn{3}{c}{TREND} &       & \multicolumn{3}{c}{ALPHA} \\
\cmidrule{2-4}\cmidrule{6-8}\cmidrule{10-12}\cmidrule{14-16}\cmidrule{18-20}
Method & Ret   & SR    & TO    &       & Ret   & SR    & TO    &       & Ret   & SR    & TO    &       & Ret   & SR    & TO    &       & Ret   & SR    & TO \\
\cmidrule{1-20}
%\cmidrule{1-4}\cmidrule{6-8}\cmidrule{10-12}\cmidrule{14-16}\cmidrule{18-20}
Original &  0.04  & 0.16  & 60\%  &       & 0.08  & 0.37  & 160\% &       & 0.17  & 0.98  & 50\%  &       & 0.25  & 0.85  & 106\% &       & 0.13  & 0.73  & 154\% \\
     Baseline & 0.01  & 0.06  & 53\%  &       & 0.13  & 0.46  & 153\% &       & 0.11  & 0.59  & 165\% &       & 0.13  & 0.46  & 91\%  &       & 0.11  & 0.67  & 126\% \\
      Transfer & \pmb{0.04} & \pmb{0.21} & \pmb{56\%} &       & \pmb{0.14} & \pmb{0.52} & \pmb{172\%} &       & \pmb{0.18} & \pmb{1.05} & \pmb{144\%} &       & 0.23  & 0.88  & 92\%  &       & 0.09  & 0.79  & 122\% \\
     Re-train & 0.03  & 0.14  & 68\%  &       & 0.09  & 0.40  & 181\% &       & 0.14  & 1.01  & 183\% &       & \pmb{0.24} & \pmb{0.90} & \pmb{88\%} &       & \pmb{0.21} & \pmb{0.81} & \pmb{98\%} \\
     Baseline + Transfer & 0.03  & 0.18  & 55\%  &       & 0.07  & 0.48  & 134\% &       & 0.15  & 0.63  & 152\% &       & 0.17  & 0.71  & 104\% &       & 0.14  & 0.75  & 124\% \\
    \bottomrule
    \end{tabular}%
    }
    \begin{tablenotes}
    \item {Note: This table presents the performance metrics of one-month long-short portfolios proposed by five different methods. To construct the portfolios, the original method applies the original trading signal, the baseline method applies the trading signal $\overline{x}_{it}$ in (\ref{IWA_signal}) with the triple-I weights generated by the RN34-D20/R20 method, the transfer method applies the trading signal $\overline{x}_{it}$ with the transferred triple-I weights from a pre-trained ResNet34 model in the RN34-D5/R5 method, the re-train method is the same as the transfer method except for re-training the ResNet34 model, and the ``baseline+transfer'' method applies the trading signal $\overline{x}_{it}$ with the averaged triple-I weights from both baseline and transfer methods. Other descriptions inherit from Table \ref{tab:daily_TWMA}.}
    \end{tablenotes}
  \label{tab:transfer}%
\end{table}%

\section{Non-technical Transfer Learning}\label{sec:news_rules}

In the previous study, we have demonstrated that the general price patterns detected by the ResNet ``trader'' can enhance the existing classical technical trading signals via the TWMA method. Nowadays, the traders in real world not only stick to the technical trading signals from the price charts, but also explore useful non-technical trading signals from other types of data source. Intuitively,
the price charts should also carry certain information of non-technical trading signals, since the non-technical trading signals affect the behaviors of traders, bringing an information flow to the price charts but in a rather intricate  manner.

To verify the above statement, we design a non-technical transfer learning procedure, based on the news signal proposed by \cite{Zhou2024Text}. This news signal is a sentiment score drawn from news articles and is categorized as a benchmark non-technical signal. To form the news signal for each stock $i$ on day $t$, we adopt the method of factor-augmented regularized model for prediction (FarmPredict) in \cite{Zhou2024Text}, based on all relevant news articles published on the \textit{Sina Finance} website (\url{https://finance.sina.com.cn}) from January 1, 2010 to May 1, 2023.
The FarmPredict method associates the response $r^{(q)}_{i, t+R}\equiv r_{i, t + R}$ with the regressor $X_{i, t}^{(q)}$ for $q = 1, \dots, Q_{i,t}$, where $X_{i, t}^{(q)}$ is generated from the $q$-th news article for stock $i$ published from the previous market close (3:00 p.m. on day $t-1$) to the current market close (3:00 p.m. on day $t$), and $Q_{i,t}$ is the number of news articles for stock $i$ published on day $t$. Using a rolling-window procedure, we train the FarmPredict model based on ten years of data in each window, and then we use this trained FarmPredict model and $X_{i,t}^{(q)}$ to compute the out-of-sample return predictions $\widehat{r}_{i,t+R}^{(q)}$ for the subsequent six months. This window rolls forward by six months, and the training and prediction processes are repeated. After the rolling-window procedure, we calculate the out-of-sample news signal for stock $i$ at day $t$:
\begin{align}\label{news_signal}
    \widetilde{r}_{i,t+R} \equiv \frac{1}{Q_{i,t}} \sum_{q = 1}^{Q_{i,t}} \widehat{r}_{i,t+R}^{(q)},
\end{align}
where $t$ spans from January 1, 2021 to May 1, 2023. Using the original news signal in (\ref{news_signal}), we apply the TWMA-RN method to obtain the TWMA-RN-based signal as done in (\ref{IWA_signal}). Similarly, we can also obtain the TWMA-CN-based and EWMA-based news signals. For each type of news signal above, we construct long-short portfolios as done for the technical trading signals.

Table \ref{tab:news} presents the out-of-sample portfolio performance across different news signals. From this table, we observe that all of considered TWMA-based news signals outperform the original news signal for proposing portfolios with one-day holding period. Specifically, the RN34-D5/R1 method achieves the best performance with the highest Sharpe ratio value of $2.04$, whereas the original method only has the Sharpe ratio value of 1.72, though it performs better than all of three EWMA methods. These findings strongly suggest that the triple-I weights derived from price trends can effectively enhance the original news signals, which are typically noisy and fragile. The reason behind is perhaps that the triple-I weights help emphasize the relevant importance of each day's news from the perspective of price trends, thereby de-noising the original news signal after the TWMA implementation.

For portfolios with one-week holding period,  we find from Table \ref{tab:news} that the values of Sharpe ratio decline dramatically, compared with those for portfolios with one-day holding period. This indicates that the long-term predictive strength of news articles on stock prices tends to be weak. Although the news signal is less effective in this scenario, some gain is still observed when using the TWMA-RN method to enhance the news signal. Notably, the RN34-D20/R5 method achieves the highest Sharpe ratio value of $0.41$ across all considered methods, whereas the original, CN-D20/R5, and EWMA94 methods earn significantly lower Sharpe ratio values of $0.15$, $0.14$, and $0.17$, respectively, representing a reduction of more than $58\%$.

Overall, our findings from Table \ref{tab:news} lead to a conclusion that the news signals can also benefit from the application of TWMA method, particularly for proposing portfolios with a short-term holding period. The non-technical transfer learning procedure above is novel in its own rights, showing us the importance of transferring knowledge from price-volume data to financial text data.

%To refine the collected dataset, we remove duplicate articles and apply random down-sampling to manage the volume of data.

\begin{table}[!ht]
  \centering
  \caption{One-day and one-week long-short portfolio performance across different news signals.}
  \scalebox{0.75}{
  \setlength{\tabcolsep}{7.9mm}{
    \begin{tabular}{ccccccccc}
    \toprule
   \multicolumn{4}{c}{Holding period: One-day}   &       & \multicolumn{4}{c}{Holding period: One-week} \\
\cmidrule{1-4}\cmidrule{6-9}    Method & Ret   & SR    & TO    &       & Method & Ret   & SR    & TO \\
\cmidrule{1-4}\cmidrule{6-9}    Original & 0.23  & 1.72  & 2741\% &       & Original & 0.03  & 0.15  & 509\% \\
\cmidrule{2-4}  \cmidrule{7-9}
& \multicolumn{3}{c}{Panel A:  TWMA-RN} & & & \multicolumn{3}{c}{Panel A:  TWMA-RN}\\
\cmidrule{2-4}  \cmidrule{7-9}
RN34-D5/R1 & \pmb{0.28}  & \pmb{2.04}  & \pmb{2989\%} &       & RN34-D5/R5 & 0.06  & 0.38  & 584\% \\
    RN34-D20/R1 & 0.22  & 1.98  & 2647\% &       & RN34-D20/R5 & \pmb{0.08}  & \pmb{0.41}  & \pmb{574\%} \\
    RN34-D60/R1 & 0.23  & 1.94  & 2304\% &       & RN34-D60/R5 & 0.03  & 0.14  & 382\% \\
\cmidrule{2-4}  \cmidrule{7-9}
& \multicolumn{3}{c}{Panel B:  TWMA-CN} & & & \multicolumn{3}{c}{Panel B:  TWMA-CN}\\
\cmidrule{2-4}  \cmidrule{7-9}
CN-D5/R1 & 0.25  & 2.00  & 2561\% &       & CN-D5/R5 & 0.02  & 0.11  & 493\% \\
    CN-D20/R1 & 0.22  & 1.92  & 2589\% &       & CN-D20/R5 & 0.02  & 0.14  & 542\% \\
    CN-D60/R1 & 0.20  & 1.90  & 2433\% &       & CN-D60/R5 & 0.01  & 0.10  & 601\% \\
\cmidrule{2-4}  \cmidrule{7-9}
& \multicolumn{3}{c}{Panel C:  EWMA} & & & \multicolumn{3}{c}{Panel C:  EWMA}\\
\cmidrule{2-4}  \cmidrule{7-9}
EWMA98 & 0.22  & 1.70  & 2639\% &       & EWMA98 & 0.03  & 0.14  & 489\% \\
    EWMA94 & 0.20  & 1.67  & 2588\% &       & EWMA94 & 0.04  & 0.17  & 516\% \\
    EWMA90 & 0.19  & 1.60  & 2582\% &       & EWMA90 & 0.04  & 0.16  & 521\% \\
    % EWMA70 & 0.19  & 1.58  & 2503\% &       & EWMA70 & 0.02  & 0.12  & 483\% \\
    % EWMA50 & 0.19  & 1.58  & 2496\% &       & EWMA50 & 0.02  & 0.10  & 427\% \\
    \bottomrule
    \end{tabular}%
    }}
    \begin{tablenotes}
    \item Note: This table presents the performance metrics of one-day and one-month long-short portfolios proposed by different methods. For the construction of portfolios, the original method uses the original news signal in (\ref{news_signal}); the RN34 methods (Panel A), CN methods (Panel B), and EWMA methods (Panel C) use the TWMA-RN-based, TWMA-CN-based, and EWMA-based news signals, respectively. Other descriptions are consistent with those in Table \ref{tab:daily_TWMA}.
    \end{tablenotes}
  \label{tab:news}%
\end{table}%

\section{Enhancement of Trading Rules}\label{trading_rules}

Technical trading rules have been employed in financial markets for over a century, with numerous studies suggesting their capability to generate predictive trading signals. Notably, the 7,846 technical rules used in \cite{sullivan1999data} and \cite{BAJGROWICZ2012473} serve as a broad universe of benchmarks for trading strategies. To leverage these benchmarks, we apply each of the 7,846 trading rules on a stock-by-stock basis to generate stock-level signals. For each rule, we then construct long-short portfolios based on these signals, where the portfolios have the holding period of either one-day or one-week. As a result, we obtain a distribution of out-of-sample annualized Sharpe ratios across different trading rule signals from the above process.

%decile spread

To examine the effectiveness of the TWMA method in the context of trading rules, we apply the TWMA method to enhance each rule's signal for constructing long-short portfolios. For clarity, we focus on the TWMA-RN-based and TWMA-CN-based trading rule signals generated from
RN34-D5/R1 and CN-D5/R1 methods (or RN34-D20/R5 and CN-D20/R5 methods), respectively, for proposing portfolios with one-day (or one-week) holding period.
As before, we can get a distribution of out-of-sample annualized Sharpe ratios across different TWMA-RN-based (or TWMA-CN-based) trading rule signals.

\begin{figure}[!h]
    \centering
    \includegraphics[width = \linewidth]{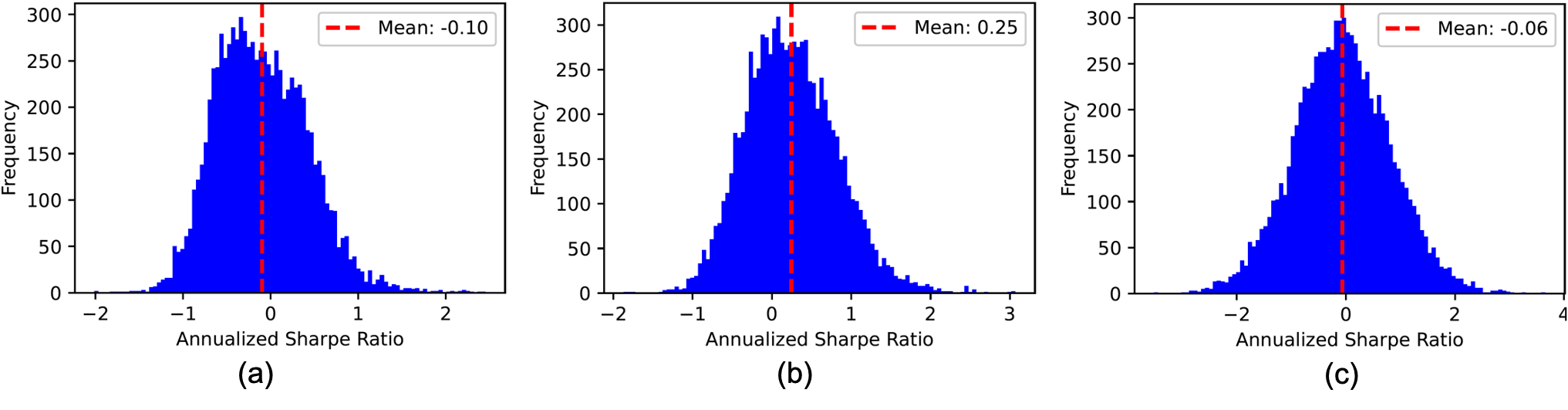}
    \caption{(Color online) One-day long-short portfolio performances for the (a) original, (b) TWMA-RN-based, and (c) TWMA-CN-based trading rule signals. The histogram in subfigure (a) illustrates the distribution of out-of-sample Sharpe ratio values for 7,846 different strategies, which are based on the original trading rule signals. The histograms in subfigures (b) and (c) depict the distributions of out-of-sampl Sharpe ratio values for 7,846 strategies, based on the TWMA-RN-based and TWMA-CN-based trading rule signals, respectively.}
    \label{fig:daily_trading_rules}
\end{figure}

Fig~\ref{fig:daily_trading_rules} exhibits three distributions of out-of-sample annualized Sharpe ratios for one-day long-short portfolios, which are proposed from the use of three different types of trading rule signal. From this figure, we find that the original trading rule signals yield a distribution of Sharpe ratio values with a sample mean of $-0.10$, while the TWMA-RN-based trading rule signals result in a distribution of Sharpe ratio values with a sample mean of $0.25$. This finding demonstrates that applying the TWMA-RN-based method to the trading rule signals can deliver a much better overall portfolio performance. In addition, when the TWMA-RN-based trading rule signals are replaced with the TWMA-CN-based trading rule signals, the sample mean of the distribution of Sharpe ratio decreases from  0.25 to -0.06. All of the aforementioned findings indicate the advantage of TWMA-RN-based trading rule signals for portfolio selections. To further check this, we calculate the rule-by-rule differenced Sharpe ratio values between any two kinds of signals. Based on the 7,846 differenced Sharpe ratio values, the classical Student's $t$ test shows that the mean of TWMA-RN-based distribution is significantly larger than that of original distribution (with a p-value of 0.027) or that of TWMA-CN-based distribution (with a p-value of 0.031). However, the result of $t$ test does not indicate that the mean of TWMA-CN-based distribution is significantly larger than that of original distribution (with a p-value of 0.142).

\begin{figure}[!h]
    \centering
    \includegraphics[width = \linewidth]{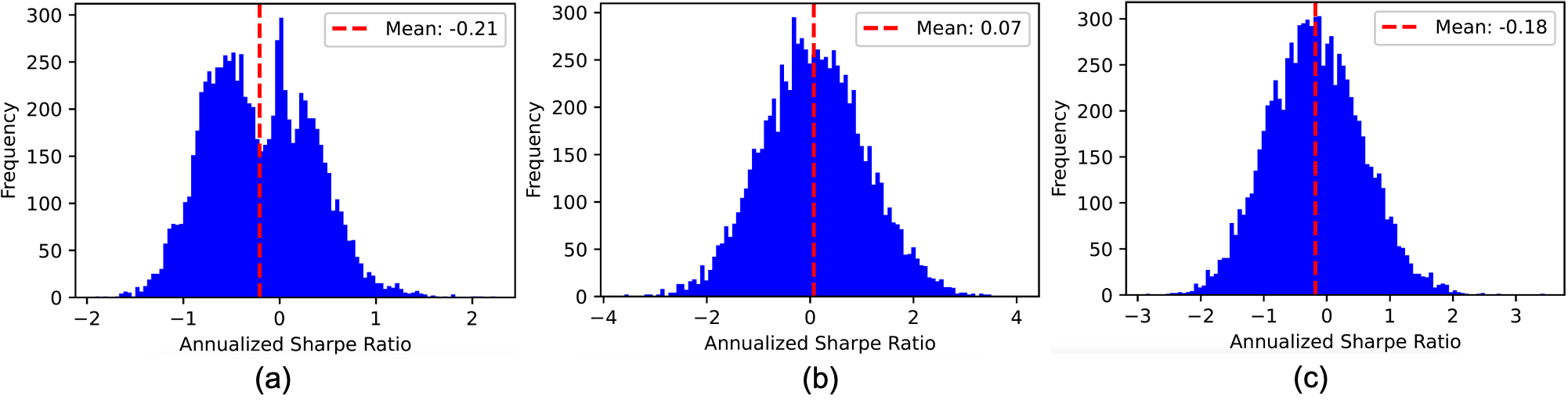}
    \caption{(Color online) One-week long-short portfolio performances for the (a) original, (b) TWMA-RN-based, and (c) TWMA-CN-based trading rule signals. Other descriptions are consistent with those in Fig~\ref{fig:daily_trading_rules}.}
    \label{fig:weekly_trading_rules}
\end{figure}

For one-week long-short portfolios, we also plot the corresponding three Sharpe ratio distributions in Fig~\ref{fig:weekly_trading_rules}. Compared with the results in Fig~\ref{fig:daily_trading_rules}, we observe a clear decline in the portfolio performance for each type of trading rule signal. Remarkably, the TWMA-RN-based distribution is still the only one having a positive sample mean. According to the results of $t$ test, we further find that the mean of TWMA-RN-based distribution is marginally larger than that of original distribution (with a p-value of 0.041) or that of TWMA-CN-based distribution (with a p-value of 0.043).
As expected, the result of $t$ test (wit a p-value of 0.208) does not support that the mean of TWMA-CN-based distribution is larger than that of original distribution.

In sum, our findings indicate the important role of TWMA method to enhance the performance of existing trading rules, particularly for targeting at portfolios with a short-term holding period.

\section{Concluding Remarks}\label{sec:conclusion}

A recent benchmark work in \cite{jiang2023re} reveals that the general price patterns detected from price charts by the CNN are informative for portfolio selections. However, they do not open up the ``black-box'' to tell the traders: What are the general price patterns detected from price charts? Furthermore, they thus can not tackle a more practically interesting question: How to make use of these detected general price patters to enhance the existing trading strategies? Using the ResNet (a much deeper neural network than the CNN), this paper deals with the first question by constructing ``downward'' and ``upward'' localization maps, which identify the important regions of price chart images for predicting the downward and upward future price movements, respectively. Based on these two localization maps, we further address the second question by designing a novel TWMA method to compute triple-I weights for enhancing the existing price trend trading signals. Our extensive empirical analysis demonstrates the usefulness of the TWMA method to enhance the commonly used price trend trading strategies in Chinese stock market, particularly for proposing portfolios with a short-term holding period. More importantly, we show the importance of triple-I weights in both time-scale transfer learning and non-technical transfer learning. Specifically, the triple-I weights trained on higher frequency data can be transferred to enhance the price trend trading signals for data at lower frequencies, and those trained on price chart images have the transferability to enhance the news trading signals from the market text data. In addition, we also illustrate that the TWMA method can enhance the overall portfolio performance
with respect to a large number of portfolios generated by 7,846 different trading rules.

The success of TWMA method provides concrete evidence that our attempts to address the aforementioned two questions have been effective.
Notably, the TWMA method keeps the interpretability of the existing technical trading signals predefined by the human, while it incorporates some general but imperceptible trading signals extracted from the price chart images by the machine.
Hence, its success encourages us to do technical analysis by combining the human domain knowledge and the machine-extracted knowledge.
Typically, the human domain knowledge is more interpretable but less objective, whereas the the machine-extracted knowledge
sacrifices certain interpretability to achieve better generalization, formalization, and transmission. Neither of them should be completely replaced by the other. In the field of technical analysis, how to propose a more effective and profitable way to utilize both kinds of knowledge is a promising future study.

\bibliographystyle{imsart-nameyear}
\bibliography{Ref}

\newpage

\setcounter{table}{0}
\setcounter{figure}{0}
\setcounter{algorithm}{0}
\counterwithin{table}{section}
\counterwithin{figure}{section}
\counterwithin{algorithm}{section}
\renewcommand{\thefigure}{A\arabic{figure}}
\renewcommand{\thealgorithm}{B\arabic{algorithm}}
\renewcommand{\thetable}{C\arabic{table}}

\appendix

\section{Architecture Details of the ResNet}\label{appen:network}

The fundamental idea of the ResNet is stacking together a sequence of convolution, activation, and pooling operations to convert a raw image into a set of predictive patterns (i.e., the extracted feature map $\vF_{i,t}^{(L)}$). To gain a better understanding of the ResNet, we provide the details of these operations in this appendix.

% These predictive patterns are then used to determine the final prediction; see \eqref{res}, \eqref{avg_pool} and \eqref{fc} for a review.
% and rescale
% rescale

\begin{figure}[!h]
    \centering
    \includegraphics[width = \textwidth]{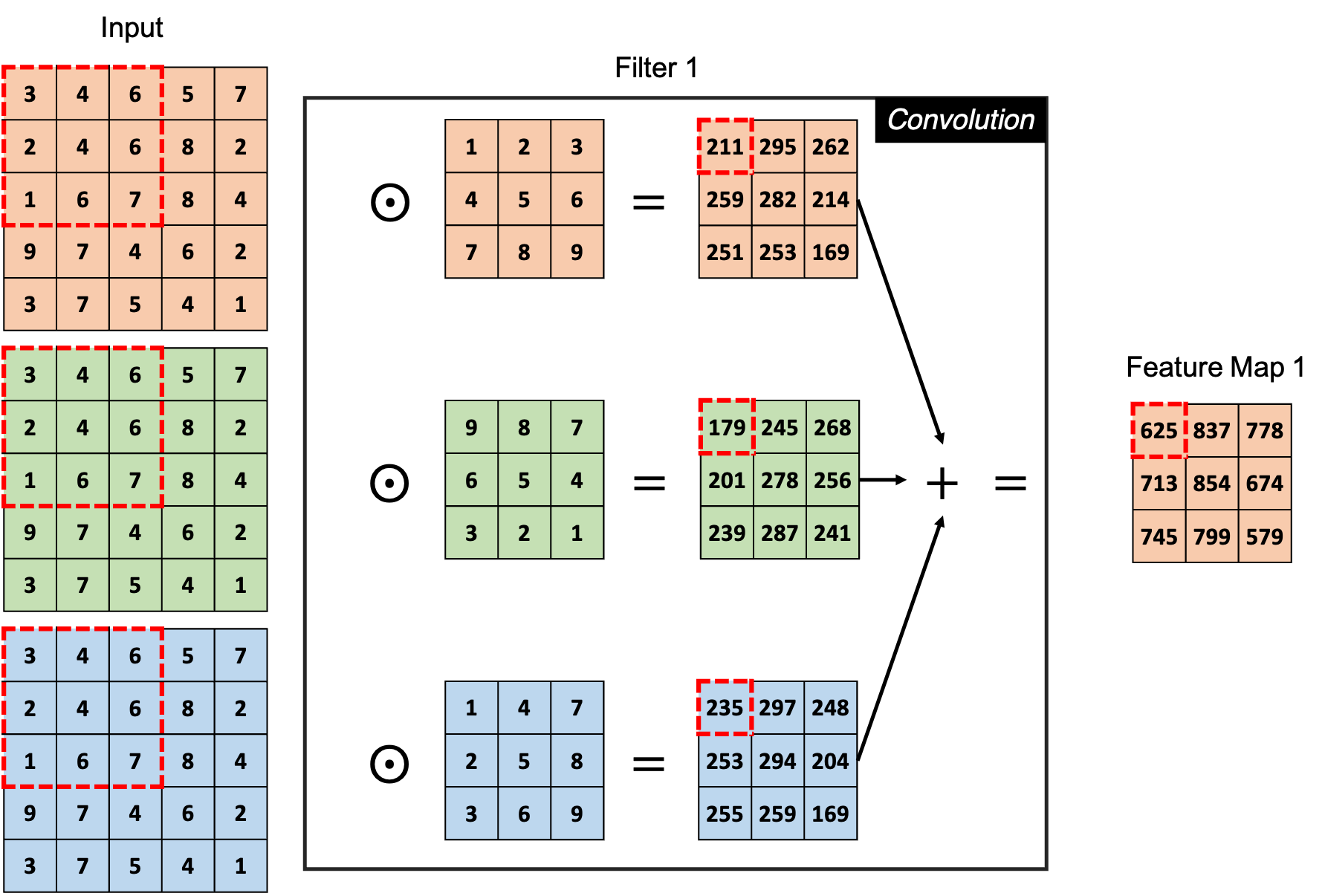}
    \caption{(Color online) Illustration of the convolution operation. The input tensor has a size of $5 \times 5 \times 3$. The matrix for each channel of input is displayed in the left column. In this example, only one $3 \times 3 \times 3$ filter, labeled as Filter 1, is considered. The filter weight matrix for each channel is shown in the middle-left column. When applying Filter 1 to the patch in the upper-left corner (indicated by a red dashed line), the convolution operation first calculates the element-wise products between the input and filter weights channel-by-channel. Then, the convolution yields a scalar for the corresponding spatial location by summing up the products across all channels. By sliding over the input with a stride of $1$, the channel-specific results are obtained, which are presented in the middle-right column. The convolution output with Filter 1 is the sum of scanned results across different channels, labeled as Feature Map 1, displayed in the right column. Intuitively, each filter corresponds to a specific feature map, and the convolution output is reassembled into a tensor with a channel number equal to the number of employed filters.}
    \label{fig:conv}
\end{figure}

The first and most crucial operation is convolution, which applies a set of learnable filters, also known as the kernels, to scan the (image) input. Compared to the input, these filters are smaller tensors, with the same number of channels (e.g., a $3 \times 3 \times 3$ tensor for colorful image input with $224 \times 224 \times 3$ shape). Specifically, each filter slides over the input with a certain stride channel-by-channel, computes the dot product between filter weights and the corresponding input patch, and calculates the sum across all channels. The convolution output, also known as the feature map, produces a summary of content in the spatially surrounding areas. See Figure \ref{fig:conv} for a visual representation of the process of convolution. Notably, the filter weights in each filter are learnable parameters (i.e., $\btheta_c^{(l)}$ in \eqref{res}), while the filter size, number of filters, and stride are user-specific hyperparameters. Evidently, the filter size and stride in the $l$-th convolution operation jointly determine the dimensions of the outputted feature map (i.e., $H^{(l)}$ and $W^{(l)}$), while the number of filters specifies the channel number of the output (i.e., $C^{(l)}$). Furthermore, the learnable filter weights within the employed filters enable the convolution to extract the most predictive aspects of the outcome of interest and blur out uninformative content in a data-driven manner.

The second essential operation, activation, is conceptually straightforward. Specifically, it is a non-linear transformation applied element-wisely to the output of a convolution operation. In this paper, the ReLU function is employed as the activation function, and mathematically it is expressed as:
\begin{align*}
    \mathrm{ReLU}(x) =
    \begin{cases}
    x, & \text{if } x > 0,
    \\
    0, & \text{otherwise}.
    \end{cases}
\end{align*}
Clearly, the ReLU function sharpens the resolution of the convolution filter output by taking the maximum value between the output of the convolution filter and zero.

\begin{figure}[!h]
    \centering
    \includegraphics[width = 0.8\linewidth]{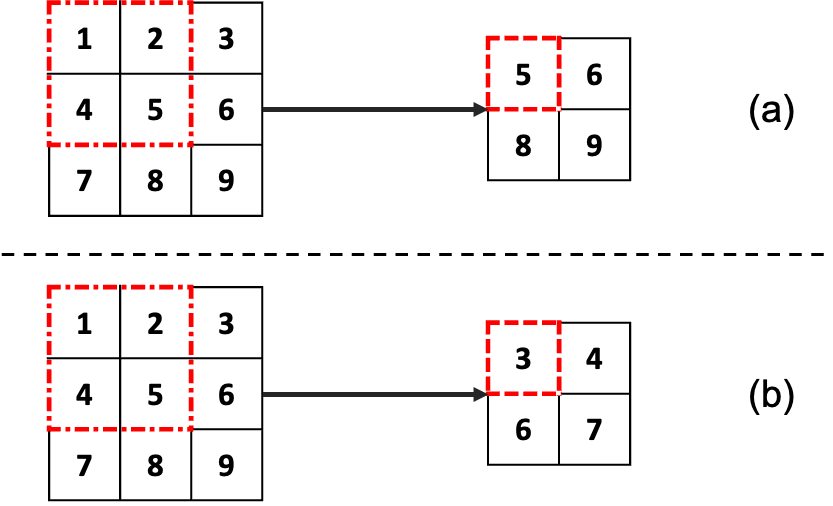}
    \caption{(Color online) Illustration of the (a) max-pooling and (b) average-pooling operations. In this example, we consider a $2 \times 2$ filter (red dashed line) for both max-pooling and average-pooling operations. The max-pooling operation scans over the input tensor with a stride of $1$ and returns the maximum value of the elements within the filter at each spatial location. In contrast, the average-pooling operation scans with the same stride but returns the average value of the elements within the filter. Both two pooling operations reduce the spatial dimension from $3 \times 3$ into $2 \times 2$.}
    \label{fig:max_pooling}
\end{figure}

The third fundamental operation is pooling. Max-pooling and average-pooling are arguably the two most widely used pooling operations; see some examples of their implementations in Fig~\ref{fig:max_pooling}. Similar to the convolution, the max-pooling operation employs a filter to scan over the input tensor channel-by-channel, and then returns the maximum value within the filter. Intuitively, max-pooling serves a two-fold purpose. First, it acts as a dimension reduction mechanism. Neighboring outputs from the convolution operation often contain similar information. If any element within the filter region is prominent, max-pooling detects it while discarding locally redundant information. Second, by selecting local maxima throughout the input, the output remains largely unaffected by minor perturbations in the input tensor. From this viewpoint, max-pooling works as a de-noising tool, enhancing robustness to local deformations. In practice, the ResNet  (illustrated in Fig~\ref{fig:network}) only applies the max-pooling operation once, with a stride of $2$ and a filter size of $3 \times 3$, after the initial $7 \times 7$ convolution.

Average-pooling also reduces the spatial dimensions of the input by scanning the input tensor with a filter. However, instead of selecting the maximum value within the filter, it computes the average of all values within the filter. The average-pooling operation is particularly useful for smoothing the inputs and preserving more of the overall information from the input, rather than focusing solely on the most prominent features. In the ResNet, average-pooling is utilized to convert the final feature map into channel-wise scores. Specifically, average-pooling with a filter size of $H^{(L)} \times W^{(L)}$ is employed to scan the $\vF_{i,t}^{(L)}$ exactly once without any movement;  see, for example, \eqref{avg_pool} for its mathematical illustration.

In addition to the aforementioned operations, a notable operation in the ResNet is the rescale operation $\mathbb{R}(\cdot)$. Note that the feature maps $\vF_{i,t}^{(l)}$ and $\vF_{i,t}^{(l - S)}$ in \eqref{res} may have different shapes, which can cause troubles for the residual learning process. To solve this problem, the rescale operation is implemented through convolution with a filter size of $1 \times 1$ and a predetermined stride, ensuring that $\mathbb{R}(\vF_{i,t}^{(l - S)})$ matches the shape of $\vF_{i,t}^{(l)}$.

Having introduced the generic architectural components of ResNet, we now specify the choices of related hyperparameters in the ResNet. The recommended filter size and number of filters for each convolution operation, as suggested by \cite{he2016resnet}, are illustrated in Figure \ref{fig:network}. The initial $7 \times 7$ convolution and max-pooling operations utilize a stride of $2$, as the input consists of sparse raw images. Following these operations, all subsequent convolution processes employ a stride of $1$. In terms of the rescale operation, a consistent stride of $2$ is implemented across all of considered ResNets.

\section{Adam Algorithm}

Algorithm \ref{alg:adam} below presents the details of the Adam algorithm to solve optimization problem \eqref{cross_entropy} for obtaining the estimated network parameters.

\begin{algorithm}[!h]
\caption{The training procedure of $\widehat{\btheta}_{D, R}$ by the Adam algorithm.}\label{alg:adam}
\begin{algorithmic}[1]
\REQUIRE ~~\\
    The sample: $\{(y_{i,t+R}, \vF_{i,t,D})\}$;\\
    The initial value of network parameters: $\btheta^{(0)}$;\\
    Hyperparameters: batch size $M$ and learning rate $\gamma$;\\
\STATE $k = 0$;
\REPEAT
\STATE $B^{(k)}$ $\leftarrow$ the minibatch of $\{(i,t): i = 1, \dots, N; t = 1, \dots, T\}$ including $M$ elements;
\STATE $\bm{g}^{(l)} \leftarrow \nabla_{\btheta} \left[ \sum_{(i,t) \in B^{(k)}}\ell(y_{i,t+R}, \vF_{i,t,D}, \btheta^{(k)}) \right]$ (Gradients of minibatch estimator);
\STATE $\btheta^{(k + 1)} \leftarrow$ update parameters using learning rate $\gamma$ and gradients $\bm{g}^{(l)}$ (Adam);
\STATE $l \leftarrow l+1$;
\UNTIL{convergence of parameters $\btheta^{(l + 1)}$};
\ENSURE ~~\\
    The value of $\btheta^{(k + 1)}$, which is taken as the estimator $\widehat{\btheta}_{D,R}$.
\end{algorithmic}
\end{algorithm}

\section{ResNet ``trader'' versus CNN ``trader''}

In Sections \ref{sec:empirical}--\ref{sec:news_rules}, we mainly focus on the empirical performance of TWMA-RN and TWMA-CN methods in enhancing trading strategies. Recall that the TWMA-RN and TWMA-CN methods rely on the triple-I weights derived from the ResNet ``trader'' and CNN ``trader'', respectively. A natural question arises: Does the better-performance of the TWMA-RN method over the TWMA-CN method result from a more sophisticated trader? To answer this question, we evaluate the economic performance of the ResNet ``trader'' and CNN ``trader'' in this appendix.

Following the methodology in \cite{jiang2023re}, we apply the predicted ``up'' probability (i.e., $\widehat{P}_{i,t+R, 1}$) calculated by each ``trader" as the trading signal to propose equal-weight long-short decile portfolios, while other details about portfolio selection remain consistent with those in Section \ref{sec:iwa_model}.

\subsection{Portfolio Performance with One-day Holding Period}

We begin with the evaluation of one-day portfolio performance. Table \ref{tab:daily_strategy} reports the one-day portfolio performance of ResNet ``trader'' and CNN ``trader'',
who use different methods (with respect to the choices of network architecture and image input) to predict up probability. As a comparison, some classical traders, using traditional technical trading signals to target one-day forward returns, are also employed as benchmarks. From Table \ref{tab:daily_strategy}, we have the following findings:

(i) For the ResNet ``trader'', regardless of network architecture and image inputs, the Sharpe ratio values increase monotonically from decile 1 (Low) to decile 10 (High). This implies that the ResNet ``trader'' is capable of distinguishing stocks with positive or negative one-day returns. The highest Sharpe ratio value, 9.48, is achieved by the long-short portfolio based on the RN34-D5/R1 method, followed by the long-short portfolios based on the RN50-D5/R1 and RN18-D5/R1 methods, which attain Sharpe ratio values of 8.27 and 8.19, respectively. These findings are consistent to those in Table \ref{tab:daily_TWMA}. For the CNN ``trader'', its Sharpe ratio values are much smaller than those obtained by the ResNet ``trader'', irrespective of the image inputs. The superiority of the ResNet ``trader'' over the CNN ``trader'' is particularly pronounced under the D5/R1 and D20/R1 settings, as the long-short portfolios selected by the RN34-D5/R1 and RN34-D20/R1 methods yield 310\% and 283\% higher values of Sharpe ratio than those selected by the CN-D5/R1 and CN-D20/R1 methods, respectively. This observation highlights the benefits of employing a deep network architecture for enhancing the predictive strength of ``trader.''

(ii) When the input is changing from D5-images to D60-images, the performance of the long-short portfolios selected by the ResNet and CNN ``traders'' becomes worse. This suggests that it is adverse to include information information from a longer time period to do shorter-term (such as one-day) return predictions.

(iii) For the classical traders, the long-short portfolios selected by the MOM, STR, and WSTR traders achieve relatively low Sharpe ratio values of 0.01, 0.18, and 0.26, respectively. In contrast, the long-short portfolios based on the TREND and ALPHA traders yield higher Sharpe ratio values of 1.54 and 2.60, respectively. Interestingly, with sufficient patterns built upon domain knowledge, the performance of the ALPHA trader is marginally superior to that of the CNN ``trader.'' However, a considerable gap remains between the portfolio performance of the ALPHA trader and the ResNet ``trader.''

\begin{table}[!h]
  \centering
  \caption{One-day portfolio performance across different traders.}
  \scalebox{0.63}{
    \begin{tabular}{ccccccccccccccccccccccccccc}
    \toprule
          & \multicolumn{26}{c}{ResNet ``trader''}   \\
%\cmidrule{2-9}\cmidrule{11-18}\cmidrule{20-27}
\cmidrule{2-27}
& \multicolumn{2}{c}{RN18-D5/R1} &       & \multicolumn{2}{c}{RN18-D20/R1} &       & \multicolumn{2}{c}{RN18-D60/R1} &       & \multicolumn{2}{c}{RN34-D5/R1} &       & \multicolumn{2}{c}{RN34-D20/R1} &       & \multicolumn{2}{c}{RN34-D60/R1} &       & \multicolumn{2}{c}{RN50-D5/R1} &       & \multicolumn{2}{c}{RN50-D20/R1} &       & \multicolumn{2}{c}{RN50-D60/R1} \\
\cmidrule{2-3}\cmidrule{5-6}\cmidrule{8-9}\cmidrule{11-12}\cmidrule{14-15}\cmidrule{17-18}\cmidrule{20-21}\cmidrule{23-24}\cmidrule{26-27}    Decile & Ret   & SR    &       & Ret   & SR    &       & Ret   & SR    &       & Ret   & SR    &       & Ret   & SR    &       & Ret   & SR    &       & Ret   & SR    &       & Ret   & SR    &       & Ret   & SR \\
\cmidrule{1-3}\cmidrule{5-6}\cmidrule{8-9}\cmidrule{11-12}\cmidrule{14-15}\cmidrule{17-18}\cmidrule{20-21}\cmidrule{23-24}\cmidrule{26-27}    Low   & -0.31  & -1.67  &       & -0.21  & -1.07  &       & -0.22  & -1.17  &       & -0.37  & -2.03  &       & -0.22  & -1.18  &       & -0.20  & -1.04  &       & -0.33  & -1.82  &       & -0.26  & -1.35  &       & -0.28  & -1.46  \\
    2     & -0.15  & -0.82  &       & -0.14  & -0.73  &       & -0.09  & -0.48  &       & -0.20  & -1.10  &       & -0.14  & -0.76  &       & -0.10  & -0.53  &       & -0.15  & -0.81  &       & -0.17  & -0.91  &       & -0.14  & -0.78  \\
    3     & -0.05  & -0.29  &       & -0.08  & -0.44  &       & -0.05  & -0.26  &       & -0.10  & -0.56  &       & -0.07  & -0.38  &       & -0.04  & -0.22  &       & -0.08  & -0.45  &       & -0.08  & -0.43  &       & -0.10  & -0.56  \\
    4     & -0.02  & -0.10  &       & -0.05  & -0.26  &       & -0.01  & -0.07  &       & -0.03  & -0.15  &       & -0.06  & -0.30  &       & -0.01  & -0.06  &       & -0.03  & -0.17  &       & -0.03  & -0.16  &       & -0.03  & -0.14  \\
    5     & 0.01  & 0.06  &       & 0.01  & 0.03  &       & 0.03  & 0.16  &       & 0.03  & 0.14  &       & 0.00  & 0.02  &       & 0.06  & 0.30  &       & 0.02  & 0.11  &       & 0.05  & 0.28  &       & 0.05  & 0.28  \\
    6     & 0.07  & 0.38  &       & 0.05  & 0.29  &       & 0.08  & 0.43  &       & 0.09  & 0.47  &       & 0.03  & 0.18  &       & 0.11  & 0.58  &       & 0.10  & 0.55  &       & 0.06  & 0.34  &       & 0.12  & 0.62  \\
    7     & 0.15  & 0.77  &       & 0.15  & 0.81  &       & 0.11  & 0.61  &       & 0.16  & 0.85  &       & 0.11  & 0.60  &       & 0.14  & 0.75  &       & 0.12  & 0.64  &       & 0.15  & 0.78  &       & 0.11  & 0.59  \\
    8     & 0.21  & 1.08  &       & 0.14  & 0.73  &       & 0.17  & 0.89  &       & 0.22  & 1.11  &       & 0.17  & 0.88  &       & 0.17  & 0.87  &       & 0.23  & 1.19  &       & 0.23  & 1.19  &       & 0.21  & 1.08  \\
    9     & 0.24  & 1.21  &       & 0.24  & 1.22  &       & 0.21  & 1.08  &       & 0.30  & 1.47  &       & 0.26  & 1.34  &       & 0.18  & 0.92  &       & 0.25  & 1.26  &       & 0.24  & 1.26  &       & 0.26  & 1.33  \\
    High  & 0.37  & 1.79  &       & 0.41  & 2.03  &       & 0.27  & 1.32  &       & 0.43  & 2.08  &       & 0.43  & 2.10  &       & 0.21  & 1.00  &       & 0.38  & 1.85  &       & 0.32  & 1.56  &       & 0.31  & 1.52  \\
    H-L   & 0.67  & 8.19  &       & 0.62  & 7.03  &       & 0.49  & 5.31  &       & 0.80  & 9.48 &       & 0.65  & 7.47  &       & 0.40  & 4.35  &       & 0.71  & 8.27  &       & 0.57  & 7.15  &       & 0.59  & 6.96  \\
    Turnover & \multicolumn{2}{c}{3665\%} &       & \multicolumn{2}{c}{3750\%} &       & \multicolumn{2}{c}{3689\%} &       & \multicolumn{2}{c}{3282\%} &       & \multicolumn{2}{c}{3398\%} &       & \multicolumn{2}{c}{3408\%} &       & \multicolumn{2}{c}{3308\%} &       & \multicolumn{2}{c}{3438\%} &       & \multicolumn{2}{c}{3389\%} \\
    \midrule
          & \multicolumn{8}{c}{CNN ``trader''}                                       &       & \multicolumn{14}{c}{Classical trader}                                                                   &       &       &  \\
\cmidrule{2-9}\cmidrule{11-24}          & \multicolumn{2}{c}{CN-D5/R1} &       & \multicolumn{2}{c}{CN-D20/R1} &       & \multicolumn{2}{c}{CN-D60/R1} &       & \multicolumn{2}{c}{MOM} &       & \multicolumn{2}{c}{STR} &       & \multicolumn{2}{c}{WSTR} &       & \multicolumn{2}{c}{TREND} &       & \multicolumn{2}{c}{ALPHA} &       &       &  \\
\cmidrule{2-3}\cmidrule{5-6}\cmidrule{8-9}\cmidrule{11-12}\cmidrule{14-15}\cmidrule{17-18}\cmidrule{20-21}\cmidrule{23-24}
Decile      & Ret   & SR    &       & Ret   & SR    &       & Ret   & SR    &       & Ret   & SR    &       & Ret   & SR    &       & Ret   & SR    &       & Ret   & SR    &       & Ret   & SR    &       &       &  \\
\cmidrule{1-3}\cmidrule{5-6}\cmidrule{8-9}\cmidrule{11-12}\cmidrule{14-15}\cmidrule{17-18}\cmidrule{20-21}\cmidrule{23-24}    Low   & -0.14  & -0.82  &       & -0.10  & -0.59  &       & -0.03  & -0.18  &       & 0.00  & -0.01  &       & -0.06  & -0.25  &       & -0.06  & -0.27  &       & -0.10  & -0.38  &       & -0.21  & -1.09  &       &       &  \\
    2     & -0.03  & -0.19  &       & -0.02  & -0.14  &       & 0.04  & 0.21  &       & 0.05  & 0.24  &       & 0.06  & 0.28  &       & -0.06  & -0.33  &       & 0.05  & 0.22  &       & -0.08  & -0.43  &       &       &  \\
    3     & 0.03  & 0.17  &       & -0.01  & -0.06  &       & 0.02  & 0.08  &       & 0.03  & 0.16  &       & 0.12  & 0.65  &       & 0.01  & 0.08  &       & 0.06  & 0.30  &       & -0.04  & -0.19  &       &       &  \\
    4     & 0.02  & 0.08  &       & 0.07  & 0.37  &       & 0.09  & 0.50  &       & 0.04  & 0.25  &       & 0.15  & 0.82  &       & 0.03  & 0.20  &       & 0.11  & 0.58  &       & -0.01  & -0.03  &       &       &  \\
    5     & 0.04  & 0.22  &       & 0.07  & 0.38  &       & 0.08  & 0.41  &       & 0.10  & 0.55  &       & 0.12  & 0.65  &       & 0.10  & 0.54  &       & 0.09  & 0.49  &       & 0.02  & 0.13  &       &       &  \\
    6     & 0.11  & 0.56  &       & 0.08  & 0.43  &       & 0.09  & 0.45  &       & 0.07  & 0.40  &       & 0.07  & 0.36  &       & 0.10  & 0.51  &       & 0.08  & 0.45  &       & 0.05  & 0.27  &       &       &  \\
    7     & 0.11  & 0.57  &       & 0.06  & 0.32  &       & 0.05  & 0.27  &       & 0.11  & 0.59  &       & 0.04  & 0.23  &       & 0.15  & 0.74  &       & 0.09  & 0.49  &       & 0.07  & 0.39  &       &       &  \\
    8     & 0.13  & 0.66  &       & 0.13  & 0.65  &       & 0.06  & 0.32  &       & 0.10  & 0.50  &       & 0.03  & 0.13  &       & 0.09  & 0.45  &       & 0.09  & 0.48  &       & 0.09  & 0.50  &       &       &  \\
    9     & 0.12  & 0.59  &       & 0.09  & 0.47  &       & 0.07  & 0.35  &       & 0.06  & 0.29  &       & 0.02  & 0.08  &       & 0.15  & 0.68  &       & -0.01  & -0.06  &       & 0.11  & 0.64  &       &       &  \\
    High  & 0.13  & 0.65  &       & 0.15  & 0.72  &       & 0.05  & 0.21  &       & 0.00  & 0.00  &       & -0.01  & -0.06  &       & 0.00  & 0.00  &       & 0.22  & 1.04  &       & 0.17  & 0.96  &       &       &  \\
    H-L   & 0.28  & 2.31  &       & 0.26  & 1.95  &       & 0.12  & 1.76  &       & 0.00  & 0.01  &       & 0.05  & 0.18  &       & 0.06  & 0.26  &       & 0.31  & 1.54  &       & 0.39  & 2.60  &       &       &  \\
    Turnover & \multicolumn{2}{c}{3112\%} &       & \multicolumn{2}{c}{3128\%} &       & \multicolumn{2}{c}{2169\%} &       & \multicolumn{2}{c}{281\%} &       & \multicolumn{2}{c}{805\%} &       & \multicolumn{2}{c}{1605\%} &       & \multicolumn{2}{c}{2422\%} &       & \multicolumn{2}{c}{3018\%} &       &       &  \\
    \bottomrule
    \end{tabular}%
    }
    \begin{tablenotes}
    \item Note: This table presents the performance metrics of one-day portfolios, which are formed based on the signals used by different traders. The metrics include the annualized excess return (Ret) and annualized Sharpe ratio (SR) for each portfolio, while the monthly turnover (TO) of the long-short (H-L) portfolio is reported separately.
    \end{tablenotes}
  \label{tab:daily_strategy}%
\end{table}%

\subsection{Portfolio Performance with Longer Holding Periods}

The monthly turnover values of the one-day long-short portfolios, selected by the ResNet ``trader'', are in excess of $3200\%$, whereas that based on the MOM signal is only $281\%$. A higher value of monthly turnover leads to significant trading costs, which can impact the practical performance of trading strategies. Below, we investigate the performance of portfolios that trade less frequently.

Table \ref{tab:weekly_strategy} reports the performance of one-week portfolios proposed by different traders, based on the predicted signals over a 5-day horizon (i.e., $R = 5$). From this table, we could draw the following conclusions:

(i) Except for the portfolios proposed by the MOM, STR, and WSTR traders, the one-week long-short portfolios have lower Sharpe ratio values and monthly turnover values than the corresponding one-day long-short portfolios in Table \ref{tab:daily_strategy}. Unfortunately, no signals maintain the monotonicity of Sharpe ratio values with respect to the decile number.

(ii) The ResNet ``trader'' continues to outperform the competitors from the viewpoint of economic performance, as evidenced by the highest Sharpe ratio value of 3.74 achieved by the RN34-D20/R5 method. With comparable values of monthly turnover, the RN34-D20/R5 method earns more than three times the Sharpe ratio value of the WSTR method and nearly double the Sharpe ratio value of the CN-D20/R5 method, which achieves the highest Sharpe ratio value among all three CNN methods.

(iii) For the classical traders, the ALPHA trader exhibits the best economic performance. However, it performs substantially worse than the CNN ``trader'' with CN-D20/R5 method. This finding highlights the limitation of patterns derived from domain knowledge, which are not robust to the forecasting horizons.

\begin{table}[!h]
  \centering
  \caption{One-week portfolio performance across different traders.}
  \scalebox{0.63}{
    \begin{tabular}{ccccccccccccccccccccccccccc}
    \toprule
          & \multicolumn{26}{c}{ResNet ``trader''}    \\
%\cmidrule{2-9}\cmidrule{11-18}\cmidrule{20-27}
\cmidrule{2-27}
& \multicolumn{2}{c}{RN18-D5/R5} &       & \multicolumn{2}{c}{RN18-D20/R5} &       & \multicolumn{2}{c}{RN18-D60/R5} &       & \multicolumn{2}{c}{RN34-D5/R5} &       & \multicolumn{2}{c}{RN34-D20/R5} &       & \multicolumn{2}{c}{RN34-D60/R5} &       & \multicolumn{2}{c}{RN50-D5/R5} &       & \multicolumn{2}{c}{RN50-D20/R5} &       & \multicolumn{2}{c}{RN50-D60/R5} \\
\cmidrule{2-3}\cmidrule{5-6}\cmidrule{8-9}\cmidrule{11-12}\cmidrule{14-15}\cmidrule{17-18}\cmidrule{20-21}\cmidrule{23-24}\cmidrule{26-27}    Decile & Ret   & SR    &       & Ret   & SR    &       & Ret   & SR    &       & Ret   & SR    &       & Ret   & SR    &       & Ret   & SR    &       & Ret   & SR    &       & Ret   & SR    &       & Ret   & SR \\
\cmidrule{1-3}\cmidrule{5-6}\cmidrule{8-9}\cmidrule{11-12}\cmidrule{14-15}\cmidrule{17-18}\cmidrule{20-21}\cmidrule{23-24}\cmidrule{26-27}    Low   & -0.09  & -0.46  &       & -0.12  & -0.56  &       & -0.07  & -0.32  &       & -0.12  & -0.60  &       & -0.10  & -0.44  &       & -0.02  & -0.10  &       & -0.08  & -0.40  &       & -0.09  & -0.43  &       & -0.02  & -0.07  \\
    2     & -0.03  & -0.13  &       & -0.07  & -0.33  &       & 0.00  & 0.02  &       & -0.03  & -0.16  &       & -0.03  & -0.13  &       & 0.00  & -0.01  &       & -0.03  & -0.14  &       & -0.07  & -0.31  &       & 0.02  & 0.10  \\
    3     & 0.06  & 0.29  &       & -0.04  & -0.19  &       & 0.01  & 0.07  &       & 0.00  & 0.02  &       & -0.04  & -0.21  &       & -0.02  & -0.11  &       & -0.01  & -0.06  &       & -0.02  & -0.09  &       & 0.02  & 0.08  \\
    4     & 0.01  & 0.03  &       & 0.02  & 0.10  &       & 0.03  & 0.15  &       & 0.04  & 0.20  &       & 0.03  & 0.16  &       & 0.02  & 0.10  &       & 0.01  & 0.06  &       & 0.02  & 0.12  &       & -0.04  & -0.17  \\
    5     & 0.06  & 0.28  &       & 0.02  & 0.10  &       & 0.04  & 0.20  &       & 0.10  & 0.51  &       & 0.05  & 0.24  &       & 0.04  & 0.20  &       & 0.04  & 0.19  &       & 0.06  & 0.30  &       & -0.06  & 0.32  \\
    6     & 0.08  & 0.38  &       & 0.07  & 0.36  &       & 0.05  & 0.25  &       & 0.06  & 0.27  &       & 0.05  & 0.26  &       & 0.05  & 0.28  &       & 0.10  & 0.47  &       & 0.09  & 0.44  &       & 0.06  & 0.28  \\
    7     & 0.07  & 0.33  &       & 0.10  & 0.50  &       & 0.12  & 0.58  &       & 0.08  & 0.40  &       & 0.10  & 0.48  &       & 0.09  & 0.45  &       & 0.10  & 0.51  &       & 0.10  & 0.47  &       & 0.08  & 0.40  \\
    8     & 0.08  & 0.41  &       & 0.11  & 0.55  &       & 0.09  & 0.48  &       & 0.09  & 0.45  &       & 0.13  & 0.63  &       & 0.09  & 0.43  &       & 0.07  & 0.35  &       & 0.08  & 0.41  &       & 0.10  & 0.49  \\
    9     & 0.11  & 0.50  &       & 0.20  & 1.02  &       & 0.10  & 0.49  &       & 0.13  & 0.59  &       & 0.17  & 0.78  &       & 0.13  & 0.64  &       & 0.16  & 0.77  &       & 0.16  & 0.78  &       & 0.07  & 0.36  \\
    High  & 0.18  & 0.82  &       & 0.23  & 1.06  &       & 0.14  & 0.64  &       & 0.17  & 0.77  &       & 0.16  & 0.77  &       & 0.14  & 0.68  &       & 0.15  & 0.69  &       & 0.18  & 0.83  &       & 0.16  & 0.77  \\
    H-L   & 0.27  & 3.26  &       & 0.35  & 3.48  &       & 0.21  & 1.92  &       & 0.29  & 3.56  &       & 0.26  & 3.74  &       & 0.17  & 1.64  &       & 0.23  & 2.62  &       & 0.27  & 2.96  &       & 0.18  & 1.85  \\
    Turnover & \multicolumn{2}{c}{705\%} &       & \multicolumn{2}{c}{659\%} &       & \multicolumn{2}{c}{677\%} &       & \multicolumn{2}{c}{706\%} &       & \multicolumn{2}{c}{683\%} &       & \multicolumn{2}{c}{669\%} &       & \multicolumn{2}{c}{709\%} &       & \multicolumn{2}{c}{679\%} &       & \multicolumn{2}{c}{686\%} \\
    \midrule
          & \multicolumn{8}{c}{CNN ``trader''}                                       &       & \multicolumn{14}{c}{Classical trader}                                                                       &       &       &  \\
\cmidrule{2-9}\cmidrule{11-24}          & \multicolumn{2}{c}{CN-D5/R5} &       & \multicolumn{2}{c}{CN-D20/R5} &       & \multicolumn{2}{c}{CN-D60/R5} &       & \multicolumn{2}{c}{MOM} &       & \multicolumn{2}{c}{STR} &       & \multicolumn{2}{c}{WSTR} &       & \multicolumn{2}{c}{TREND} &       & \multicolumn{2}{c}{ALPHA} &       &       &  \\
\cmidrule{2-3}\cmidrule{5-6}\cmidrule{8-9}\cmidrule{11-12}\cmidrule{14-15}\cmidrule{17-18}\cmidrule{20-21}\cmidrule{23-24}
Decile     & Ret   & SR    &       & Ret   & SR    &       & Ret   & SR    &       & Ret   & SR    &       & Ret   & SR    &       & Ret   & SR    &       & Ret   & SR    &       & Ret   & SR    &       &       &  \\
\cmidrule{1-3}\cmidrule{5-6}\cmidrule{8-9}\cmidrule{11-12}\cmidrule{14-15}\cmidrule{17-18}\cmidrule{20-21}\cmidrule{23-24}    Low   & -0.05  & -0.24  &       & -0.08  & -0.41  &       & -0.02  & -0.08  &       & 0.00  & -0.01  &       & -0.10  & -0.40  &       & -0.11  & -0.49  &       & -0.05  & -0.21  &       & -0.16  & -0.76  &       &       &  \\
    2     & -0.01  & -0.07  &       & -0.02  & -0.10  &       & 0.01  & 0.05  &       & 0.03  & 0.12  &       & 0.02  & 0.13  &       & 0.02  & 0.12  &       & -0.01  & -0.07  &       & -0.08  & -0.39  &       &       &  \\
    3     & 0.03  & 0.16  &       & 0.04  & 0.18  &       & 0.00  & 0.00  &       & 0.05  & 0.22  &       & 0.15  & 0.77  &       & 0.06  & 0.34  &       & -0.01  & -0.07  &       & -0.04  & -0.17  &       &       &  \\
    4     & 0.03  & 0.17  &       & 0.04  & 0.21  &       & 0.02  & 0.09  &       & 0.06  & 0.28  &       & 0.10  & 0.54  &       & 0.10  & 0.50  &       & 0.01  & 0.04  &       & -0.01  & -0.04  &       &       &  \\
    5     & 0.11  & 0.53  &       & 0.02  & 0.12  &       & 0.06  & 0.31  &       & 0.10  & 0.53  &       & 0.09  & 0.49  &       & 0.10  & 0.52  &       & 0.07  & 0.32  &       & 0.02  & 0.08  &       &       &  \\
    6     & 0.01  & 0.20  &       & 0.08  & 0.41  &       & 0.08  & 0.42  &       & 0.08  & 0.44  &       & 0.08  & 0.37  &       & 0.10  & 0.41  &       & 0.07  & 0.38  &       & 0.06  & 0.29  &       &       &  \\
    7     & 0.08  & 0.36  &       & 0.09  & 0.47  &       & 0.07  & 0.32  &       & 0.09  & 0.46  &       & 0.06  & 0.26  &       & 0.09  & 0.41  &       & 0.11  & 0.53  &       & 0.06  & 0.30  &       &       &  \\
    8     & 0.09  & 0.43  &       & 0.12  & 0.59  &       & 0.10  & 0.48  &       & 0.11  & 0.54  &       & 0.06  & 0.26  &       & 0.07  & 0.30  &       & 0.16  & 0.78  &       & 0.05  & 0.25  &       &       &  \\
    9     & 0.12  & 0.54  &       & 0.12  & 0.57  &       & 0.09  & 0.40  &       & 0.03  & 0.13  &       & 0.03  & 0.12  &       & 0.07  & 0.29  &       & 0.13  & 0.58  &       & 0.09  & 0.48  &       &       &  \\
    High  & 0.11  & 0.33  &       & 0.11  & 0.51  &       & 0.11  & 0.51  &       & 0.02  & 0.06  &       & 0.02  & 0.08  &       & 0.02  & 0.07  &       & 0.22  & 0.86  &       & 0.12  & 0.64  &       &       &  \\
    H-L   & 0.12  & 1.12  &       & 0.19  & 2.08  &       & 0.12  & 1.19  &       & 0.02  & 0.07  &       & 0.12  & 0.46  &       & 0.12  & 0.63  &       & 0.27  & 1.11  &       & 0.28  & 1.77  &       &       &  \\
    Turnover & \multicolumn{2}{c}{705\%} &       & \multicolumn{2}{c}{677\%} &       & \multicolumn{2}{c}{669\%} &       & \multicolumn{2}{c}{129\%} &       & \multicolumn{2}{c}{357\%} &       & \multicolumn{2}{c}{682\%} &       & \multicolumn{2}{c}{406\%} &       & \multicolumn{2}{c}{582\%} &       &       &  \\
    \bottomrule
    \end{tabular}%
    }
    \begin{tablenotes}
    \item Note: This table presents the performance metrics of one-week portfolios, which are formed based on the signals used by different traders. Other descriptions are the same as those in Table \ref{tab:daily_strategy}.
    \end{tablenotes}
  \label{tab:weekly_strategy}%
\end{table}%

As supplementary information, Table \ref{tab:monthly_strategy} presents the one-month portfolio performance across different traders,
based on the predicted signals over a 20-day horizon (i.e., $R = 20$). With a longer holding period, the dominance of image-based signals used by the ResNet and CNN ``traders'' diminishes, while the signals used by the traditional traders begin to exhibit improved performance.

Overall, the ResNet ``trader'' outperforms the CNN ``trader'' from multiple perspectives, providing a rationale for why the TWMA-RN method enhances trading strategies more effectively than the TWMA-CN method. Note that although the portfolios proposed by the ResNet ``trader'' have larger values of Sharpe ratio than those proposed by TWMA-RN method, they are much less profitable due to their much larger values of monthly turnover.

% \section{Long-term Portfolio Performance}

\begin{table}[!h]
  \centering
  \caption{One-month portfolio performance across different traders.}
  \scalebox{0.63}{
    \begin{tabular}{ccccccccccccccccccccccccccc}
    \toprule
          & \multicolumn{26}{c}{ResNet ``trader''}     \\
%\cmidrule{2-9}\cmidrule{11-18}\cmidrule{20-27}
\cmidrule{2-27}
& \multicolumn{2}{c}{RN18-D5/R20} &       & \multicolumn{2}{c}{RN18-D20/R20} &       & \multicolumn{2}{c}{RN18-D60/R20} &       & \multicolumn{2}{c}{RN34-D5/R20} &       & \multicolumn{2}{c}{RN34-D20/R20} &       & \multicolumn{2}{c}{RN34-D60/R20} &       & \multicolumn{2}{c}{RN50-D5/R20} &       & \multicolumn{2}{c}{RN50-D20/R20} &       & \multicolumn{2}{c}{RN50-D60/R20} \\
\cmidrule{2-3}\cmidrule{5-6}\cmidrule{8-9}\cmidrule{11-12}\cmidrule{14-15}\cmidrule{17-18}\cmidrule{20-21}\cmidrule{23-24}\cmidrule{26-27}   Decile & Ret   & SR    &       & Ret   & SR    &       & Ret   & SR    &       & Ret   & SR    &       & Ret   & SR    &       & Ret   & SR    &       & Ret   & SR    &       & Ret   & SR    &       & Ret   & SR \\
\cmidrule{1-3}\cmidrule{5-6}\cmidrule{8-9}\cmidrule{11-12}\cmidrule{14-15}\cmidrule{17-18}\cmidrule{20-21}\cmidrule{23-24} \cmidrule{26-27}   Low   & 0.07  & 0.40  &       & 0.00  & 0.00  &       & -0.04  & -0.20  &       & 0.04  & 0.20  &       & -0.03  & -0.14  &       & 0.00  & -0.02  &       & 0.07  & 0.37  &       & -0.05  & -0.26  &       & -0.03  & -0.15  \\
    2     & 0.02  & 0.09  &       & 0.05  & 0.29  &       & -0.02  & -0.11  &       & 0.03  & 0.15  &       & 0.03  & 0.16  &       & 0.00  & 0.00  &       & 0.04  & 0.21  &       & 0.00  & 0.03  &       & 0.04  & 0.19  \\
    3     & 0.01  & 0.07  &       & 0.02  & 0.09  &       & 0.01  & 0.06  &       & 0.05  & 0.30  &       & 0.02  & 0.13  &       & 0.03  & 0.17  &       & 0.04  & 0.22  &       & 0.02  & 0.10  &       & 0.04  & 0.22  \\
    4     & 0.02  & 0.10  &       & 0.01  & 0.04  &       & 0.05  & 0.25  &       & 0.02  & 0.14  &       & 0.03  & 0.15  &       & 0.04  & 0.22  &       & 0.02  & 0.09  &       & 0.03  & 0.20  &       & 0.03  & 0.14  \\
    5     & 0.01  & 0.07  &       & 0.04  & 0.24  &       & 0.04  & 0.21  &       & 0.07  & 0.41  &       & 0.03  & 0.17  &       & 0.00  & -0.02  &       & 0.01  & 0.08  &       & 0.02  & 0.08  &       & 0.04  & 0.24  \\
    6     & 0.08  & 0.42  &       & 0.02  & 0.14  &       & 0.08  & 0.43  &       & 0.05  & 0.26  &       & 0.06  & 0.32  &       & 0.05  & 0.28  &       & 0.05  & 0.28  &       & 0.07  & 0.35  &       & 0.02  & 0.12  \\
    7     & 0.05  & 0.30  &       & 0.07  & 0.38  &       & 0.05  & 0.25  &       & 0.02  & 0.10  &       & 0.05  & 0.27  &       & 0.06  & 0.34  &       & 0.06  & 0.31  &       & 0.09  & 0.47  &       & 0.07  & 0.38  \\
    8     & 0.04  & 0.24  &       & 0.06  & 0.33  &       & 0.07  & 0.39  &       & 0.04  & 0.24  &       & 0.06  & 0.33  &       & 0.06  & 0.35  &       & 0.05  & 0.27  &       & 0.06  & 0.32  &       & 0.05  & 0.30  \\
    9     & 0.05  & 0.31  &       & 0.09  & 0.49  &       & 0.08  & 0.45  &       & 0.02  & 0.11  &       & 0.10  & 0.58  &       & 0.07  & 0.38  &       & 0.03  & 0.18  &       & 0.06  & 0.34  &       & 0.07  & 0.42  \\
    High  & 0.07  & 0.36  &       & 0.08  & 0.42  &       & 0.12  & 0.63  &       & 0.08  & 0.46  &       & 0.08  & 0.43  &       & 0.12  & 0.57  &       & 0.06  & 0.33  &       & 0.08  & 0.41  &       & 0.11  & 0.55  \\
    H-L   & 0.00  & -0.06  &       & 0.08  & 0.71  &       & 0.16  & 0.94  &       & 0.05  & 0.51  &       & 0.10  & 0.79  &       & 0.12  & 0.63  &       & -0.01  & -0.13  &       & 0.13  & 0.72  &       & 0.14  & 0.85  \\
    Turnover & \multicolumn{2}{c}{171\%} &       & \multicolumn{2}{c}{170\%} &       & \multicolumn{2}{c}{163\%} &       & \multicolumn{2}{c}{169\%} &       & \multicolumn{2}{c}{171\%} &       & \multicolumn{2}{c}{166\%} &       & \multicolumn{2}{c}{169\%} &       & \multicolumn{2}{c}{170\%} &       & \multicolumn{2}{c}{167\%} \\
    \midrule
          & \multicolumn{8}{c}{CNN ``trader''}                                       &       & \multicolumn{14}{c}{Classical trader}                                                                       &       &       &  \\
\cmidrule{2-9}  \cmidrule{11-24}        & \multicolumn{2}{c}{CN-D5/R20} &       & \multicolumn{2}{c}{CN-D20/R20} &       & \multicolumn{2}{c}{CN-D60/R20} &       & \multicolumn{2}{c}{MOM} &       & \multicolumn{2}{c}{STR} &       & \multicolumn{2}{c}{WSTR} &       & \multicolumn{2}{c}{TREND} &       & \multicolumn{2}{c}{ALPHA} &       &       &  \\
\cmidrule{2-3}\cmidrule{5-6}\cmidrule{8-9}\cmidrule{11-12}\cmidrule{14-15}\cmidrule{17-18}\cmidrule{20-21}\cmidrule{23-24}
Decile     & Ret   & SR    &       & Ret   & SR    &       & Ret   & SR    &       & Ret   & SR    &       & Ret   & SR    &       & Ret   & SR    &       & Ret   & SR    &       & Ret   & SR    &       &       &  \\
\cmidrule{1-3}\cmidrule{5-6}\cmidrule{8-9}\cmidrule{11-12}\cmidrule{14-15}\cmidrule{17-18}\cmidrule{20-21}\cmidrule{23-24}    Low   & 0.03  & 0.14  &       & 0.00  & 0.00  &       & -0.01  & -0.07  &       & -0.02  & -0.10  &       & -0.06  & -0.31  &       & -0.14  & -1.12  &       & -0.10  & -0.50  &       & -0.09  & -0.45  &       &       &  \\
    2     & 0.03  & 0.16  &       & 0.03  & 0.16  &       & 0.02  & 0.10  &       & 0.02  & 0.09  &       & -0.02  & -0.10  &       & -0.05  & -0.35  &       & -0.02  & -0.10  &       & -0.03  & -0.15  &       &       &  \\
    3     & 0.08  & 0.40  &       & 0.07  & 0.37  &       & 0.07  & 0.43  &       & 0.09  & 0.51  &       & 0.09  & 0.53  &       & 0.04  & 0.31  &       & 0.01  & 0.06  &       & -0.01  & -0.03  &       &       &  \\
    4     & 0.05  & 0.29  &       & 0.02  & 0.14  &       & 0.02  & 0.10  &       & 0.05  & 0.28  &       & 0.08  & 0.52  &       & 0.07  & 0.59  &       & 0.02  & 0.10  &       & 0.01  & 0.04  &       &       &  \\
    5     & 0.03  & 0.19  &       & 0.04  & 0.20  &       & 0.06  & 0.36  &       & 0.08  & 0.48  &       & 0.09  & 0.50  &       & 0.09  & 0.65  &       & 0.05  & 0.32  &       & 0.02  & 0.10  &       &       &  \\
    6     & 0.04  & 0.26  &       & 0.07  & 0.35  &       & 0.04  & 0.22  &       & 0.05  & 0.31  &       & 0.07  & 0.38  &       & 0.10  & 0.61  &       & 0.10  & 0.58  &       & 0.02  & 0.14  &       &       &  \\
    7     & 0.01  & 0.08  &       & 0.05  & 0.28  &       & 0.06  & 0.34  &       & 0.10  & 0.56  &       & 0.08  & 0.43  &       & 0.09  & 0.49  &       & 0.13  & 0.71  &       & 0.05  & 0.26  &       &       &  \\
    8     & 0.04  & 0.23  &       & 0.07  & 0.38  &       & 0.05  & 0.28  &       & 0.07  & 0.36  &       & 0.04  & 0.18  &       & 0.06  & 0.34  &       & 0.11  & 0.56  &       & 0.04  & 0.23  &       &       &  \\
    9     & 0.05  & 0.27  &       & 0.06  & 0.30  &       & 0.05  & 0.28  &       & 0.02  & 0.09  &       & 0.05  & 0.22  &       & 0.10  & 0.53  &       & 0.13  & 0.61  &       & 0.05  & 0.30  &       &       &  \\
    High  & 0.06  & 0.34  &       & 0.04  & 0.21  &       & 0.07  & 0.36  &       & 0.02  & 0.08  &       & 0.02  & 0.10  &       & 0.03  & 0.16  &       & 0.14  & 0.51  &       & 0.05  & 0.30  &       &       &  \\
    H-L   & 0.04  & 0.54  &       & 0.04  & 0.42  &       & 0.09  & 0.78  &       & 0.04  & 0.16  &       & 0.08  & 0.37  &       & 0.17  & 0.98  &       & 0.25  & 0.85  &       & 0.13  & 0.73  &       &       &  \\
    Turnover & \multicolumn{2}{c}{170\%} &       & \multicolumn{2}{c}{167\%} &       & \multicolumn{2}{c}{165\%} &       & \multicolumn{2}{c}{60\%} &       & \multicolumn{2}{c}{160\%} &       & \multicolumn{2}{c}{50\%} &       & \multicolumn{2}{c}{106\%} &       & \multicolumn{2}{c}{154\%} &       &       &  \\
    \bottomrule
    \end{tabular}%
     }
     \begin{tablenotes}
    \item Note: This table presents the performance metrics of one-month portfolios, which are formed based on the signals used by different traders. Other descriptions are the same as those in Table \ref{tab:daily_strategy}.
    \end{tablenotes}
  \label{tab:monthly_strategy}%
\end{table}%

\end{document}